\documentclass{emulateapj}
\usepackage{graphicx}
\usepackage{amsmath}
\usepackage{amssymb} 
\usepackage{subfigure}
\usepackage{float}

\begin{document}

\newcommand{\gcm}{\text{ g cm}^{-3}}
\newcommand{\mm}{\ \mu \text{m}}
\newcommand{\cm}{\ \text{cm}}
\newcommand{\mg}{\ \mu \text{G}} 
\newcommand{\kms}{\ \text{km s}^{-1}}
\newcommand{\AU}{\ \text{AU}}
\newcommand{\V}{\ \text{Volt}}
\newcommand{\Gyr}{\ \text{Gyr}} 
\newcommand{\Myr}{\ \text{Myr}}
\newcommand{\pc}{\ \text{pc}}


\def\etal{et al.\ \rm}
\def\ba{\begin{eqnarray}}
\def\ea{\end{eqnarray}}

\title{Probing Oort Cloud and Local Interstellar Medium Properties via Dust Produced 
in Cometary Collisions}

\author{Alex R. Howe\altaffilmark{1} and 
Roman R. Rafikov\altaffilmark{1}}
\altaffiltext{1}{Department of Astrophysical Sciences, 
Princeton University, Ivy Lane, Princeton, NJ 08540, USA; 
arhowe@astro.princeton.edu, rrr@astro.princeton.edu}


\begin{abstract}
{The Oort Cloud remains one of the most poorly explored regions of the Solar System. We propose that its properties can be constrained by studying a population of dust grains produced in collisions of comets in the outer Solar System. We explore the dynamics of ${\rm \mu m}$-size grains outside the heliosphere (beyond $\sim 250$ AU), which are affected predominantly by the magnetic field of the interstellar medium (ISM) flow past the Sun. We derive analytic models for the production and motion of small particles as a function of their birth location in the Cloud and calculate the particle flux and velocity distribution in the inner Solar System. These models are verified by direct numerical simulations. We show that grains originating in the Oort Cloud have a unique distribution of arrival directions, which should easily distinguish them from both interplanetary and interstellar dust populations. We also demonstrate that the distribution of particle arrival velocities is uniquely determined by the mass distribution and dust production rate in the Cloud. Cometary collisions within the Cloud produce a flux of ${\rm \mu}$m-size grains in the inner Solar System of up to several m$^{-2}$ yr$^{-1}$. The next-generation dust detectors may be sensitive enough to detect and constrain this dust population, which will illuminate us about the Oort Cloud's properties. We also show that the recently-detected mysterious population of large (${\rm \mu}$m-size) unbound particles, which seems to arrive with the ISM flow, is unlikely to be generated by collisions of comets in the Oort Cloud.} 
\end{abstract}



\section{Introduction.}  
\label{sect:intro}


The Outer Solar System (hereafter OSS) is a home of the Oort Cloud of comets \citep{Oort} --- an extended reservoir of primitive objects that provides a source of the long-period comets observed in the Inner Solar System (hereafter ISS). We define OSS as the region of space outside the helioshere --- the inner part of the Solar System within about 250 AU from the Sun \citep{Richardson,McComas} affected by the Solar wind. The population of comets in the outer region of the Oort Cloud ($>10^4$ AU) is estimated at $(5-10)\times 10^{11}$ objects larger than 2.3 km, based on observations of long-period comets in the ISS \citep{Dones}. The nature of the inner parts of the Oort Cloud ($<10^4$ AU), while potentially more populated, is not as well known because objects on Keplerian orbits originating in this region are not likely to reach inside the orbit of Jupiter because of the weaker influence of the galactic tide \citep{Dones}.

Some of the Oort Cloud's properties have been inferred from numerical studies of its formation and dynamics. The first numerical study of this sort conducted by \cite{Duncan} produced an Oort Cloud with a sharp inner edge at $r_{in} \sim 3,000$ AU and a density profile of $r^{-3.5}$ between $3\times 10^3$ AU and $5\times 10^4$ AU. A later work by \cite{Dones} also finds $r_{in} \approx 3\times 10^3$ AU, but an outer edge as far as  $2\times 10^5$ AU. The observed distribution of long-period comets indicates that their source population extends inward to about 3,000 AU but does not provide information about nearer distances \citep{Kaib09}.

These simulations model the formation of the Solar System in isolation. However, the Sun is thought to have formed in a cluster environment, which could have resulted in stronger external tides, truncating the Cloud \citep{Brasser}. Focusing on this possibility, \cite{Kaib08} find that Oort Cloud objects could form with semi-major axes as small as 100 AU. The discovery of Kuiper Belt objects like (90377) Sedna, which has a perihelion of 76 AU (too far to be affected by planetary perturbations) and an aphelion of $\sim 900$ AU has led to speculation of an ``inner Oort Cloud'' that extends significantly closer than 3,000 AU \citep{Brown}.

An added complication is that the inner part of the Oort Cloud is not necessarily isotropic, but may form a thick disk concentrated near the ecliptic. \cite{Dones} find that the isotropization radius may be anywhere from 200 AU to $2\times 10^4$ AU, depending on the density of the parent cluster in which the Sun has formed. In any case, it may be several times $r_{in}$.


\subsection{Cometary dust as a probe of the Oort Cloud.}
\label{sect:dust_OC}

Given these uncertainties about the Cloud properties, any additional independent ways of gaining information about its structure and characteristics become extremely valuable. Here we explore the possibility of probing the Cloud structure indirectly via the {\it measurements of properties of dust particles} that ought to be produced in the Oort Cloud in {\it collisions of its constituent comets}. 

Previously, comets have been invoked to explain the origin of dust particles observed in the ISS in two possible contexts. First, Jupiter-family comets --- relatively short-period comets thought to originate in the Kuiper Belt \citep{Duncan88} --- are considered as one of the sources for the zodiacal cloud particles in the ISS \citep{Nesvorny_JFC}. This source is very different from the Oort Cloud population, which we consider in this work. Second, disruptions of the long-period Oort Cloud comets {\it in the ISS} due to non-gravitational forces were explored by \citet{Nesvorny_JFC} as the source of relatively large ($\sim 100\mu$m) Earth-crossing dust particles. Our work is distinct from \citep{Nesvorny_JFC} in that it focuses on the {\it in situ} production of dust particles in cometary collisions {\it in the Oort Cloud itself}, i.e. far from the Sun. 

The production rate of cometary dust in the Oort Cloud, as we show later, is a sensitive function of both the spatial distribution of comets inside the Cloud and their physical properties. If these particles end up in the ISS, the measurement of their characteristics can {\it inform us about the Cloud's properties}. It is obvious that inferring the latter via observations of cometary dust is a two-step exercise --- one needs to understand both the {\it production} of cometary dust inside the Cloud and its subsequent {\it transport} into the ISS. In this work we address both aspects of the problem.

The dynamics of small particles in the ISS is largely determined by the effects of the Solar wind, radiation forces and planetary perturbations \citep{Moro-Martin}. However, in the OSS the trajectories of cometary dust particles must be dominated by a different set of forces (besides the Solar gravity). Since the Solar wind does not penetrate beyond the heliosphere, the Oort Cloud at distances $\gtrsim 10^3$ AU is directly exposed to an unperturbed flow of interstellar medium (ISM). The latter arises because the Sun moves at a velocity of $v_w\approx 26$ km s$^{-1}$ through the warm phase of the ISM, which is characterized by a gas number density $n({\rm H}^0)=0.2$ cm$^{-3}$, temperature $T_g=6300$ K, and ionization fraction $\chi \approx 0.25$ (Frisch \etal 2009). The strength and orientation of the magnetic field carried with the wind are rather uncertain, but a typical estimate is $B\sim 3-5\,\mu$G (Opher \etal 2009). A summary of the ISM wind parameters from the literature can be found in Table \ref{ISMmodels}. The values adopted in this study, which we base on the most recent measurements, are listed in the first row.

\begin{table*}[h]
\begin{center}
\caption{Various published measurements and models of the properties of the ISM and IMF in the vicinity of the Solar System.}
\begin{tabular}{|l|l|l|l|l|l|}
\hline
v$_\infty$ (km s$^{-1}$) & $n_e$ (cm$^{-3}$) & T (K) & IMF (${\rm \mu}$G) & $\theta_{wB}$\tablenotemark{a} (deg) & Reference \\
\hline
23.2                  & 0.07                   & 6300                  & 2.2       & 42                  & Our Adopted Values  \\
26.4\tablenotemark{b} & 0.095\tablenotemark{b} & 6400\tablenotemark{b} & 2.2       & 42                  & \cite{Ben-Jaffel}   \\
23.2(0.3)             & 0.07\tablenotemark{b}  & 6300(390)             & $\ge$3.0  & 45\tablenotemark{b} & \cite{McComas}      \\
22.8                  &                        & 6                  &           &                     & \cite{Bzowski}      \\
26.3                  & 0.06                   & 6300                  & 2.7       &                     & \cite{Frisch2}      \\
26.3\tablenotemark{b} &                        & 6300\tablenotemark{b} & 3.3       & 46                  & \cite{Frisch10}     \\
23.2(0.3)\tablenotemark{b} &                   &                       & $\sim$3.0 & 76                  & \cite{Frisch12}     \\
26.4\tablenotemark{b} & 0.06\tablenotemark{b}  & 6527\tablenotemark{b} & $\ge$3.0  &                     & \cite{Pogorelov}    \\
26.3(0.4)             &                        & 6300(340)             &           &                     & \cite{Witte}        \\
26.24(0.45)           &                        & 6306(390)             &           &                     & \cite{Mobius}       \\
25\tablenotemark{b}   & 0.07\tablenotemark{b}  & 6306(390)             & 1.8\tablenotemark{b} & 40(10)   & \cite{Ratkiewicz}   \\
26.4\tablenotemark{b} & 0.04-0.11\tablenotemark{b} & 6400\tablenotemark{b} & 3.8   & 30                  & \cite{Ratkiewicz1}  \\
26.4\tablenotemark{b} & 0.08                   & 6400\tablenotemark{b} & 3.0(1.0)  & 40                  & \cite{Grygorczuk}   \\
                      &                        &                       & 3.7-5.5   & 20-30               & \cite{Opher}        \\
26.4\tablenotemark{b} & 0.06\tablenotemark{b}  & 6527\tablenotemark{b} & $\ge$4.0  & 30                  & \cite{Pogorelov09}  \\
                      & 0.08                   &                       & 3.0       &                     & \cite{Heerikhuisen} \\
26.4\tablenotemark{b} & 0.09                   & 6527\tablenotemark{b} & 2.0-3.0   &                     & \cite{Heerikhuisen11} \\
26.4\tablenotemark{b} & 0.08                   &                       & 4.4       & 20                  & \cite{Chalov}       \\
26.4\tablenotemark{b} & 0.095\tablenotemark{b} & 6400\tablenotemark{b} & 2.4(0.3)  & 39(9)               & \cite{Strumik}      \\
\hline
\end{tabular}
\label{ISMmodels}
\tablenotetext{1}{The angle between ${\bf v}_w$ and {\bf B}.}
\tablenotetext{2}{Assumed.}
\end{center}
\end{table*}

Dust grains in the OSS can be affected by this wind through a variety of forces (Belyaev \& Rafikov 2010; hereafter BR10), reviewed in \S \ref{perturbations}. Some of the dust produced in the Oort Cloud may be accelerated onto hyperbolic trajectories by the ISM wind and swept into the ISS, where it may be detected by particle detectors onboard different spacecraft and by interplanetary radars (see Table \ref{dust}). The hyperbolic trajectories of these dust grains make them easy to distinguish from bound particles closer to the Sun. Possible confusion with interstellar dust grains originating in outflows from asymptotic giant branch stars and debris disks around young stars \citep{Murray} can be avoided because the latter arrive into the ISS with the ISM wind velocity, parallel to the direction\footnote{Notably, \citet{Kruger07} claim that the arrival direction of these particles has shifted by 30$^\circ$ in the past 20 years.} of the Solar motion \citep{Landgraf}, while the cometary dust has a rather distinct velocity distribution, as we show in this work, see \S \ref{sect:small_dyn}. Compositional differences between the cometary grains (icy) and ISM grains (silicates or carbon compounds) can also help discriminate between the two unbound dust populations.

A number of experiments have explored the properties of particles of different sizes arriving in the ISS on hyperbolic trajectories. These have included spacecraft particle detectors, radar observations, and optical meteor surveys, each of which probes different sizes of particles. 

For a wide range of particle sizes, the flux of interstellar grains of mass $m_g$ in the vicinity of Earth's orbit is given by $dn/d\ln m\approx 400\left(6\times 10^{-13} {\rm g}/m_g\right)^{1.1}$m$^{-2}$ yr$^{-1}$ \citep{Murray}. Dust detectors onboard {\it Galileo} and {\it Ulysses} have observed particles at 2.5-5 AU and found the flux in that region to be about one half that in the vicinity of Earth's orbit \citep{Kruger07,Altobelli05}. However, the flux of small particles (radii $<0.2-0.4$ ${\rm \mu m}$, depending on the phase of the Solar cycle) is higher at these distances because there is less filtering by Solar wind and radiation pressure further out. Based on typical ISM properties, \cite{Grun} expect typical interstellar dust grains to have $r_g \sim 0.05-0.2 {\rm \mu m}$ and  $m_g \sim 10^{-14}$ g, with a mass flux of $1.5\times 10^{-9}$ g m$^{-2}$ yr$^{-1}$. This corresponds to a particle flux of $5\times 10^3$ m$^{-2}$ yr$^{-1}$. This is much higher than the 30 m$^{-2}$ yr$^{-1}$ flux of $m_g>6\times 10^{-13}$ g particles detected by {\it Ulysses} \citep{Murray}, which can be understood as the result of small dust filtration by the Solar wind \citep{Grun}. Quite unexpectedly, it was also found \citep{Landgraf} that the interstellar dust entering the Solar System includes a population of $\gtrsim 1\mu$m particles, larger than expected to be present in the ISM. The origin of these large particles arriving with the ISM wind is not well understood \citep{Draineconundrum} and is explored in \S \ref{sect:appl2}.

Finally, optical meteor surveys suggest that larger ``interstellar'' meteors ($m_g \sim 10^{-4}$ g, $r_g \sim 300$ ${\rm \mu m}$) may comprise up to 0.01\% of the meteors striking Earth \citep{Hawkes}. A summary of the results of these studies is shown in Table \ref{dust}, where the particle radii are estimated assuming a bulk density of $\sim 3$ g cm$^{-3}$.

\begin{table*}[h]
\begin{center}
\caption{Flux of particles on hyperbolic trajectories observed in the 
vicinity of Earth's orbit.}
\begin{tabular}{|l|l|l|l|l|}
\hline
Mass (g)               & Radius (${\rm \mu m}$) & Flux (m$^{-2}$ 
yr$^{-1}$) & Facility                 & Reference        \\
\hline
$5\times 10^{-14}-1\times 10^{-12}$ & 0.16-0.43 & 790(160) & {\it 
Cassini} & \cite{Altobelli} \\
$6\times 10^{-13}-1\times 10^{-11}$ & 0.36-0.93 & 360 & {\it Ulysses, 
Galileo} & \cite{Murray}    \\
$1\times 10^{-12}-1\times 10^{-11}$\tablenotemark{a} & 0.43-0.93 & 
82(9) & {\it Helios}\tablenotemark{b} & \cite{Altobelli06} \\
$>1\times 10^{-11}$                 & 0.93      & 30 & {\it Ulysses} & 
\cite{Landgraf}  \\
$4\times 10^{-9}$                   & 6.8       & 
0.019\tablenotemark{c} & Aricebo                & \cite{Meisel}    \\
$2\times 10^{-7}$                   & 25        & $(1.3\times 
10^{-4})$\tablenotemark{c} & AMOR & \cite{Baggaley}  \\
$>1\times 10^{-5}$                  & 93        & $(1.9\times 
10^{-7})$\tablenotemark{d} & CMOR & \cite{Weryk}     \\
\hline
\end{tabular}
\label{dust}
\tablenotetext{1}{Estimated from grain charges.}
\tablenotetext{2}{Observed inward to 0.3 AU.}
\tablenotetext{3}{$m(dn/dm)$ at the given mass, roughly equal within 20\% to 
the flux of all particles larger than this size under the $m=-1.1$ power law.}
\tablenotetext{4}{Flux of particles measured to be unbound by 2$\sigma$.}
\end{center}
\end{table*}

In this work we explore the dynamics of dust grains produced in the Oort Cloud by analytic theory and numerical simulations, building off the work of BR10. Based on the observations above and the dynamical arguments below, particles with a size of 1-3 ${\rm \mu m}$ (a regime accessible to spacecraft particle detectors) appear to be of most interest. Our calculations provide us with a test of techniques to determine the inner edge of the Oort Cloud and the spatial distribution of dust production inside the Cloud. In addition to learning about the local ISM properties and the Oort Cloud itself, this will help to assess the possibility (first mentioned in Frisch et al. 1999) that anomalous large ``interstellar'' dust grains observed by {\it Ulysses}, {\it Galileo}, and {\it Cassini} \citep{Landgraf} have an Oort Cloud origin.

The structure of this paper is as follows. We start by describing the processes responsible for transporting cometary dust grains from the Oort Cloud into the ISS. In \S \ref{perturbations} we discuss the forces affecting dust grains in the OSS, while in \S \ref{sect:small_dyn} we investigate analytically the grain motion arising from these forces, including observables in the ISS. We present our numerical simulations in \S \ref{numerical}, calculate the production rate of cometary dust in the Oort Cloud in \S \ref{sect:dust_prod}, and compare the simulations against our analytical results in \S \ref{sect:disc}. We discuss the limitations of our calculations in \S \ref{sect:cav} and, finally, applications of our results in Section \ref{sect:concl}.


\section{Forces determining grain dynamics}
\label{perturbations}


BR10 have analyzed forces acting on dust grains of different sizes in
the OSS as the Sun traverses the different phases of the ISM.
For the warm phase through which the Sun is passing right now, they 
find that the effects of radiation forces and the galactic tide are 
negligible compared with the other forces that we discuss next. We start 
with electromagnetic forces which owe their existence to the fact 
that (1) the ISM is magnetized (see Table \ref{ISMmodels}) and (2) dust 
grains both in the ISM and the ISS are generally 
expected to be charged to a potential of a few volts \citep{Drainebook}.

In the Solar reference frame, Solar motion relative to the ISM 
induces an electric field ${\bf E}={\bf v}_w \times {\bf B}/c$, 
while the magnetic field strength stays essentially the same 
as in the ISM frame. Letting $U$ be the grain potential, the electric 
${\bf F_{\text{E}}}$ and magnetic ${\bf F_{\text{B}}}$ forces on a 
grain are 
\ba
{\bf F}_{\text{E}} &=& -\frac{U r_g}{c} {\bf v}_w \times {\bf B},
\label{eforce}
\\
{\bf F}_{\text{B}} &=& \frac{U r_g}{c} {\bf v} \times {\bf B},
\label{bforce}
\ea
where ${\bf v}_w$ and ${\bf v}$ are the wind and grain 
velocities in the Solar frame, respectively. The grain charge $q=Ur_g$ is 
taken to be constant. Clearly,
the magnetic force cannot be neglected compared with ${\bf F}_{\text{E}}$
if the grain gets accelerated to a speed $v_g$ comparable to $v_w$.

In addition to electromagnetic forces, grains in orbit around the Sun
experience gas and Coulomb drag due to the ISM. Since the
mean free path of gas molecules and ions in the ISM is much larger than
$r_g$, the drag force on a spherical grain under the
assumption of sticking or specular reflection is given by
\citep{Baines,Drainedrag}
\begin{eqnarray}
{\bf F}_{\text{drag}} = F_{\text{drag}}\frac{{\bf v}_w-{\bf v}}
{|{\bf v}_w-{\bf v}|},~~~~
F_{\text{drag}}=\pi r_g^2 P {\cal F}, 
\label{drag}
\end{eqnarray}
where
\begin{eqnarray}
P &\equiv & n k T 
\end{eqnarray}
is the full ISM pressure ($n$ is the particle number density, $T$ is
the temperature) and ${\cal F}$ is a dimensionless parameter, the value 
of which can be found in e.g. \citet{Belyaev}. 

The full ISM-related force acting on a particle is
\ba
{\bf F}&=&{\bf F_{\text{B}}}+{\bf F}_{\text{E}}+{\bf F}_{\text{drag}} 
\nonumber \\
&=& \frac{U r_g}{c} {\bf v} \times {\bf B}-\frac{U r_g}{c} {\bf v}_w 
\times {\bf B}+ F_{\text{drag}}\frac{{\bf v}_w-{\bf v}}
{|{\bf v}_w-{\bf v}|}
\label{eq:full}
\\
&=& S\left[\frac{{\bf v} \times {\bf B}}{v_w B}-
\frac{{\bf v}_w \times {\bf B}}{v_w B}+
C_d\frac{{\bf v}_w-{\bf v}_w}
{|{\bf v}_w-{\bf v}_w|}\right],
\label{eq:full_scaled}
\ea
where we defined parameters 
\ba
S &=& \frac{U r_g v_w B}{c}=1.2\times 10^{-16}\mbox{dyne}
~U_1 r_{g,1} v_{w,23}B_5,
\label{eq:S}
\\
C_d&=&\frac{\pi r_g P c}{U v_w B}{\cal F}=6\times 10^{-3}
\frac{r_{g,1} P_1 {\cal F}_{20}}{U_1 v_{w,23} B_5}.
\label{eq:S_d}
\ea
Here the normalized variables are the grain potential 
$U_1\equiv U/(1\,\text{V})$, the grain size $r_{g,1}\equiv r_g/(1\,\mu$m), 
the strength of the ISM magnetic field $B_5\equiv B/(5\,\mu$G), 
the ISM wind speed $v_{w,23}\equiv v_w/(23$ km s$^{-1}$), the parameter 
${\cal F}_{20} = {\cal F}/20$, and the typical ISM pressure 
$P_1 = nkT/(\text{eV} \cm^{-3})$. 

Since $C_d\ll 1$ for micron size 
grains, we will subsequently neglect the drag force for such particles 
compared with electromagnetic forces. Thus, it is fair to approximate 
micron size particle motion as being affected only by electromagnetic 
and gravitational forces, which we do from now on. 

Defining $\theta_{wB}$ to be the angle between the magnetic field and the
wind direction, $\cos\theta_{wB}\equiv {\bf B}\cdot {\bf v}_w/(Bv_w)$, the strength of 
the induction electric force $F_{\text{E}} = Ur_gv_w B\sin\theta_{wB}/c$ 
relative to the gravitational force $F_g=GM_\odot m_g/r^2$ 
($m_g$ is the grain mass, $r$ is the distance from the Sun) is
\ba
\frac{F_{\text{E}}}{F_g} \approx 50 \, r_3^2\,
U_1\, r_{g,1}^{-2}\, B_5\, v_{w,23}\,\rho_1^{-1} \sin\theta_{wB}.
\label{eq:Eforce_rel}
\ea
Here bulk density is $\rho_1\equiv \rho/(1\,\text{g cm}^{-3})$, and we
assume spherical particles: $m_g=(4\pi/3)\rho_g r_g^3$. A notable 
feature of the induction electric force $F_{\text{E}}$ is that its magnitude
and direction are independent of either the grain's speed or its 
location with respect to the Sun, provided that the magnetic field
is homogeneous on scales comparable to the size of the Solar System. 

Because $F_{\text{E}}\propto r_g$ while $F_g\propto r_g^3$, 
the induction electric force exceeds Solar gravity for particles smaller than some critical radius 
$r_{g,\text{min}}$. Grains with $r_g<r_{g,\text{min}}$ are
swept up in the ISM flow and ejected from the Solar System. 
BR10 show that an estimate of $r_{g,\text{min}}$ good to within about
 a factor of two can be obtained by setting $F_E/F_g = 0.25$, in
which case we find (neglecting the $\theta_{wB}$-dependence)
\ba
\label{rgminE}
r_{g,\text{min}} & \approx &
\frac{3}{\pi}r\left(\frac{U Bv_w \sin\theta_{wB}}{GM_\odot c\rho}
\right)^{1/2}
\\
& \approx &
14 \mm \, r_3\, \left(\frac{U_1 B_5 v_{w,23} \sin\theta_{wB}}{\rho_1}
\right)^{1/2}.
\nonumber
\ea

Equations (\ref{eq:Eforce_rel})-(\ref{rgminE}) show that for ${\rm \mu m}$-size 
particles released at the initial distance $r_0\gtrsim 10^3$ AU Solar 
gravity is just a small perturbation initially, so that we can neglect 
$F_g$ compared with the electromagnetic forces $F_{\text{E}}$ and $F_{\text{B}}$. 
Then in the frame of the wind, a newly created (e.g. in collisions 
of bigger grains) particle moves with speed close to $-{\bf v}_w$ 
since the local Keplerian velocity in the OSS is
\ba
v_K\approx 1~\mbox{km s}^{-1}r_3^{-1/2}\ll v_w.
\label{eq:v_K}
\ea
This causes gyration of the particles in the wind frame, while
in the Solar frame particles additionally experience 
an ${\bf E}\times {\bf B}$ drift with speed ${\bf v}_w$. If the
angle $\theta_{wB}$ between ${v}_w$ and ${\bf B}$ is not small, 
particles get accelerated to a velocity $\sim v_w$ in the
Solar frame on a length scale of order the Larmor radius
\ba
R_L&=& \frac{m_g v_w c}{U B r_g \sin\theta_{wB}} 
\nonumber \\
&\approx&
1.2\times 10^4 \AU \, \frac{r_{g,1}^2 \rho_1\, v_{w,23}}{U_1\, B_5\, \sin\theta_{wB}}.
\label{eq:R_L}
\ea
This estimate shows that in the OSS, ${\rm \mu m}$-size cometary grains 
can be easily entrained in the ISM wind since their $R_L$
is smaller than the Cloud size of $\sim 10^5$ AU.


\section{Dynamics of small (${\rm \mu m}$-sized) grains.}
\label{sect:small_dyn}


We now provide an approximate analytical description
of the grain motion in the OSS affected by the ISM wind.
Based on the results of the previous section we write
the equation of motion for micron size particles as
\ba
m_g\frac{d^2 {\bf r}}{dt^2}=\frac{q}{c}
\left({\bf v}-{\bf v}_w\right)\times {\bf B}-
\frac{GM_\odot m_g}{r^3}(1-\beta){\bf r},
\label{eq:eq_mot}
\ea
where $\beta$ is the ratio of the radiation pressure on the grain to the gravitational force. \citet{Burns} find that $\beta < 0.3$ and is size-dependent for icy grains larger than 1 ${\rm \mu m}$. Because the effective gravity is a small effect in most of the OSS, we omit the radiation pressure term for simplicity. The first term in Eq. (13) accounts for the Lorentz and induction electric forces, see equations (\ref{eforce})-(\ref{bforce}), while the last term describes Solar gravity.
 
A typical small particle produced inside the Oort Cloud
is accelerated by electromagnetic forces to a speed 
of order $v_w$ when it reaches the vicinity of the 
Sun. As equation (\ref{eq:Eforce_rel}) indicates, at 
separations of $\lesssim 100$ AU 
Solar gravity becomes stronger than the electromagnetic 
forces (i.e. $F_E/F_g\lesssim 1$).
However, by that time particle is moving so fast that it gets 
strongly deflected by the Sun (by $\sim 90^\circ$) only if it 
approaches the Solar sphere of gravitational influence 
of radius $r_{\rm grav}\sim GM_\odot/v_w^2$, which is of order 
one AU. This is much smaller than both the initial particle 
separation from the Sun ${\bf r}_0$ and the particle Larmor 
radius $R_L$. As a result, far from the Sun, on scales $\gg 100$ AU, 
we may neglect the Solar gravity and drop the 
last term in equation (\ref{eq:eq_mot}). Its effect on the 
near-Sun particle motion will be included later in 
\S \ref{sect:part_flux}.

The solution of the equation of motion (\ref{eq:eq_mot}) 
with initial conditions ${\bf r}(0)={\bf r}_0$, 
${\bf v}(0)={\bf v}_0$ and only the electromagnetic forces 
included is 
\ba
{\bf r}(t)&=&{\bf r}_0+\left({\bf v}_{0,\parallel}+
{\bf v}_{w,\perp}\right)t
\nonumber \\
&+&R_L\left[\sin\Omega_L t~{\bf k}
+\left(\cos\Omega_L t-1\right){\bf m}
\right],
\label{eq:r_t}
\ea
where the subscripts $\parallel$ and $\perp$ refer to the components
along and perpendicular to the magnetic field ${\bf B}=B{\bf b}$, 
$|{\bf b}|=1$, i.e. $a_\parallel \equiv {\bf a}\cdot {\bf b}$, 
$a_\perp\equiv {\bf a}-{\bf b}\left({\bf a}\cdot {\bf b}\right)$ 
for an arbitrary vector ${\bf a}$. We have also defined 
orthogonal unit vectors 
\ba
{\bf k}=\frac{{\bf v}_{0,\perp}^{\rm cm}}{v_{0,\perp}^{\rm cm}},~~~
{\bf m}=\frac{{\bf b}\times{\bf v}_{0,\perp}^{\rm cm}}
{v_{0,\perp}^{\rm cm}},
\label{eq:ort_defs}
\ea
where
\ba
{\bf v}_0^{\rm cm}={\bf v}_0-{\bf v}_w,
\label{eq:v_cm}
\ea
is the initial particle velocity at $t=0$ in the frame 
{\it co-moving with the ISM wind}. Also, $R_L=v_{0,\perp}^{\rm cm}/\Omega_L$, 
where 
\ba
\Omega_L=\frac{qB}{m_gc}\approx 4\times 10^{-4}\mbox{yr}^{-1} 
 U_1B_5r_{g,1}^{-2}\rho_1^{-1}
\label{eq:Larmor}
\ea
is the Larmor frequency. 

\begin{figure}[htp]			
\begin{center}
\includegraphics[bb = 240 400 592 718,clip,width=\columnwidth]{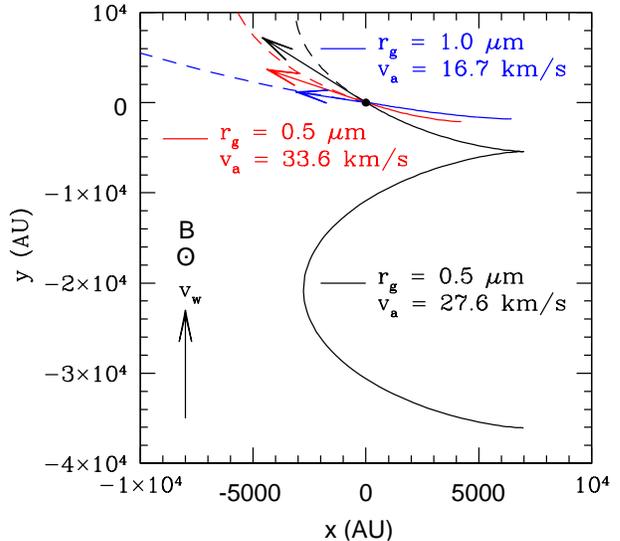}
\end{center}
\caption{Characteristic trajectories of particles originating in the Oort Cloud, moving under the influence of the ISM wind and passing through the inner Solar System (black dot). Motion in the plane perpendicular to the ${\bf B}$ field ($z$-direction) is shown, with ISM wind velocity ${\bf v}_w$ pointing in the $y$-direction (an angle $\theta_{wB} = 90^\circ$ between them is assumed). To better show the trochoidal (and nearly-cycloidal) shape of the orbits, trajectories of two smaller particles (0.5 ${\rm \mu m})$ are shown starting at different positions. The trajectory of a bigger, 1.0 ${\rm \mu m})$, particle is also displayed. Arrows near the Solar location show particle approach speeds, indicated by numbers.}
\label{orbits}
\end{figure}

Equation (\ref{eq:r_t}) can be interpreted as stating that 
the starting location ${\bf r}_0$ of all particles passing close 
to the Sun and created in the Cloud time $t_0$ ago, i.e. the ones 
for which ${\bf r}(t_0)=0$, are uniquely related to $t_0$ via
\ba
{\bf r}_0&=&-\left({\bf v}_{0,\parallel}+
{\bf v}_{w,\perp}\right)t_0
\nonumber \\
&-&R_L\left[
\sin\Omega_L t_0~{\bf k}+\left(\cos\Omega_L t_0-1\right){\bf m}
\right].
\label{eq:r0_t}
\ea

This equation shows that particle motion can be separated into two components: uniform motion along the field with the initial particle velocity $v_{0,\parallel} {\bf b}$ and a {\it trochoidal} motion in the plane perpendicular to ${\bf B}$, which is a superposition of particle gyration around ${\bf b}$ and the uniform motion of the guiding center with velocity ${\bf v}_{w,\perp}$. Sample orbits for two different particle sizes are shown in Figure \ref{orbits}. These orbits reveal a degeneracy in reconstructing particle orbits: grains approaching the Sun with the same velocity ${\bf v}_w$ may do so after different numbers of trochoid loops, obscuring their true birth distance. However, this degeneracy has a spatial period of $\sim 2\pi R_L$, which is $\gtrsim 10^5$ AU, see Eq. (\ref{eq:R_L}) for the particle sizes we consider ($1-3 {\rm \mu m}$). Cloud density is low at these distances so only a few particles will be subject to this degeneracy.

Differentiating equation (\ref{eq:r_t}) with respect to time 
we obtain the velocity of a particle created at time $t_0$ ago in
the Solar reference frame:
\ba
{\bf v}(t_0)&=&{\bf v}_{0,\parallel}+
{\bf v}_{w,\perp}
\nonumber \\
&+&v_{0,\perp}^{\rm cm}\left[
\cos(\Omega_L t_0){\bf k}-\sin(\Omega_L t_0){\bf m}\right].
\label{eq:v_t}
\ea
Its absolute value is given by 
\ba
v^2(t_0)&=&v_{0,\parallel}^2+v_{w,\perp}^2+
(v_{0,\perp}^{{\rm cm})^2}
\nonumber \\
&+&2v_{0,\perp}^{\rm cm} v_{w,\perp}
\cos\left(\Omega t_0+\varphi\right),~~~
\\
\cos\varphi&=&\frac{{\bf v}_{0,\perp}^{\rm cm}\cdot{\bf v}_{w,\perp}}
{v_{0,\perp}^{\rm cm} v_{w,\perp}}.
\label{eq:v_abs}
\ea
According to this expression the extremal
values of $v^2$ are
\ba
v_{0}^2,~~~
v_{0,\parallel}^2+\left(v_{0,\perp}-2v_{w,\perp}\right)^2.
\label{eq:min_max}
\ea

We define the grain {\it approach} velocity to the inner 
Solar System ${\bf v}_a$ as the velocity with which it would pass 
through ${\bf r}=0$ provided that the gravitational 
influence of the Sun is neglected. Equations 
(\ref{eq:v_t})-(\ref{eq:v_abs}) hold for ${\bf v}_a$ with the
understanding that ${\bf v}_a$ is a function of $t_0$, which 
is directly related to the birth location ${\bf r}_0$ via equation 
(\ref{eq:r0_t}). Thus, $t_0$ is a parameter
providing a connection between ${\bf v}_a$ and ${\bf r}_0$.


\subsection{Cold limit.}
\label{sect:cold}

Formulae describing particle motion simplify in the 
case of ${\bf v}_0=0$, i.e. when grains are born at rest with 
respect to the Sun. This assumption may also be applied
in the ``cold'' limit, when the particles are born with 
speed $v_0\ll v_w$. Given that $v_0$ is expected to be of 
order the Keplerian speed $v_K\ll v_w$ at the particle 
birth location, the cold limit should be applicable
throughout the whole Oort Cloud.

From equation (\ref{eq:r0_t}), it is clear that in the cold 
limit particle trajectories become cycloids. From equation 
(\ref{eq:v_abs}) one also finds that in this limit the 
grain approach velocity is 
\ba
v_a(t_0)=2v_{w,\perp}\left|\sin\frac{\Omega_L t_0}{2}\right|,
\label{eq:v_cold}
\ea
where $v_{w,\perp}=v_w\sin\theta_{wB}$ and $\theta_{wB}$ is the angle 
between ${\bf B}$ and ${\bf v}_w$, and 
$\cos\theta_{wB}=({\bf b}\cdot {\bf v}_w)/v_w$.
Thus, the lowest velocity with which a particle can approach 
the Sun (neglecting the gravitational acceleration due to the 
latter) in this limit is $v_a=0$, and is realized in particular 
for $t_0=0$, i.e. for particles born very close to the Sun, at 
$r_0\ll R_L$ (given by equation (\ref{eq:R_L}); again, 
neglecting Solar gravity). The 
highest particle velocity in the cold limit is $2v_{w,\perp}$.

Equation (\ref{eq:r0_t}) allows us to express the birth 
distance from the Sun in the cold limit as
\ba
&r_0^2& = 2R_L^2\left(1-\cos\Omega_L t_0\right)-
2R_L v_{w,\perp}t_0\sin\Omega_L t_0+v_{w,\perp}^2t_0^2
\nonumber\\
&=& R_L^2\left[
\left(\frac{v_a}{v_{w,\perp}}\right)^2
-4\frac{v_a}{v_{w,\perp}}
\sqrt{1-\left(\frac{v_a}{2v_{w,\perp}}\right)^2}
\arcsin\frac{v_a}{2v_{w,\perp}}\right]
\nonumber \\
&+&R_L^2\left[4\left(\arcsin
\frac{v_a}{2v_{w,\perp}}\right)^2\right],
\label{eq:r_0_2}
\ea
where in deriving the last line we have used the
relation between $t_0$ and $v_a$ from equation 
(\ref{eq:v_cold}). Equation (\ref{eq:r_0_2}) provides
a unique relation between the starting distance $r_0$ 
and the approach velocity $v_a$ near the Sun.


\subsection{Particle velocity distribution in the inner 
Solar System.}
\label{sect:part_flux}

Next we compute the distribution of particle approach 
velocity $F(v_a)$ in the Solar vicinity. We define 
it such that $F(v_a)dv_a$ is the number flux of 
particles crossing a unit surface area normal 
to ${\bf v}_a$ per unit time with velocity between $v_a$ and 
$v_a+dv_a$. It is implicitly assumed that $F(v_a)$ is also
a function of particle size, since $R_L$ and $\Omega_L$ 
depend on $r_g$. We would like to relate $F(v_a)$ to the 
production rate of particles ${\dot Q}({\bf r}_0)$ at their 
birth location ${\bf r}_0(v_a)$, i.e. the number of particles of a given size
created per unit volume and unit time at the distance 
${\bf r}_0$ from the Sun. 

According to equation (\ref{eq:v_abs}), particles 
separated by the velocity increment $dv_a$ correspond
to different ``lookback time'' moments $t_0$, separated by 
$dt_0=2v_a|\partial v_a^2/\partial t_0|^{-1} dv_a$. This difference
in $t_0$ translates to a difference in ${\bf r}_0$ given by
\ba
d{\bf r}_0=\frac{\partial {\bf r}_0}{\partial t_0}dt=-{\bf v}_a
dt_0, 
\label{eq:dr0}
\ea
where the last equality used the fact that 
$\partial {\bf r}_0/\partial t_0=-{\bf v}_a$ as follows 
from equations (\ref{eq:r0_t}) and (\ref{eq:v_t}).

\begin{figure}[htp]			
\begin{center}
\includegraphics[width=\columnwidth]{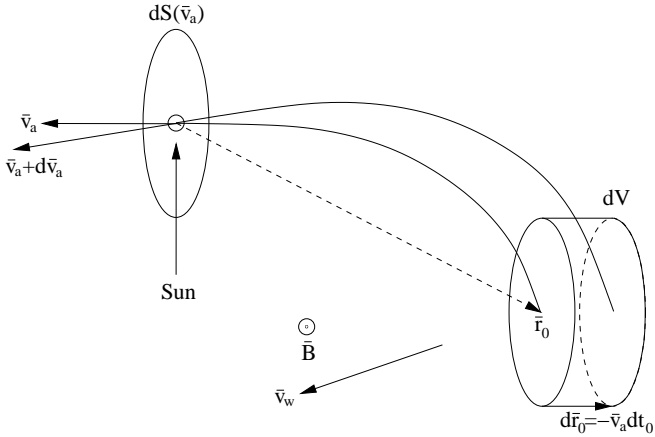} 
\end{center}
\caption{Particle trajectories originating from a volume element $dV$, located at a position $r_0$, pass through a surface element $dS$ normal to particle approach velocity ${\bf v}_a$ in the inner Solar System with uniquely determined velocities between $v_a$ and $v_a+dv_a$.}
\label{flux}
\end{figure}

Let us now consider an area element $dS({\bf v}_a)$ normal to 
the velocity ${\bf v}_a$ and run particle trajectories 
corresponding to this velocity through the edges of
the element, see Figure \ref{flux} for illustration. 
This defines a tube filled with 
particle orbits passing through $dS$ with velocity 
${\bf v}_a$. We then do the same for particle trajectories 
corresponding to velocity ${\bf v}_a+d{\bf v}_a$, which 
defines another tube filled with a different collection of 
orbits. It is easy to see that particles crossing $dS$ with 
velocities in the interval $(v_a,v_a+dv_a)$ per unit time 
$dt$ (not the lookback time increment $dt_0$!) are the 
particles created in time $dt$ inside the small cylindrical 
volume $dV$ between the faces of the two tubes on their other 
ends. The faces of this volume are oriented in the same way
as $dS({\bf v}_a)$ since the faces pass through 
${\bf r}_0(t_0(v_a))$ and ${\bf r}_0(t_0(v_a+dv_a))$, and we 
showed before that $\partial {\bf r}_0/\partial t_0=-{\bf v}_a$. 

The volume enclosed between these faces is  
\ba
dV=|d{\bf r}_0|dS=v_a dt_0 dS=
2v_a^2\left|\frac{\partial v_a^2}{\partial t_0}\right|^{-1} dv_a dS
\label{eq:dV}
\ea
and there are $dN={\dot Q}({\bf r}_0) dV dt$ particles that are created 
in this volume per unit time $dt$. All these particles get 
swept by the ISM wind and after traveling for time $t_0$
end up crossing the area element $dS$ in Solar vicinity
with speeds between $v_a$ and $v_a+dv_a$. 
Using definition of $F(v_a)$ and equation (\ref{eq:dV}) one
then finds that
\ba
F(v_a)=2v_a^2\left|\frac{\partial v_a^2}{\partial t_0}\right|^{-1}
{\dot Q}({\bf r}_0(v_a)).
\label{eq:flux}
\ea
Note that this formula does not assume a ``cold limit'' ---
in the case of non-zero ${\bf v}_0$ both $v_a$ and ${\bf r}_0$ 
are also functions of ${\bf v}_0$ and the particle flux becomes 
a function of ${\bf v}_0$ as well --- $F(v_a,{\bf v}_0)$. 
If particles are produced with a distribution of ${\bf v}_0$ 
at each point in space, then one needs to convolve 
$F(v_a,{\bf v}_0)$ with a distribution of ${\bf v}_0$ 
to get the full flux as a function of $v_a$.  

However, for our current application, it is sufficient to consider 
the cold limit. In that case using equations (\ref{eq:v_cold}) 
and (\ref{eq:flux}) we arrive at the following expression for 
the approach velocity distribution:
\ba
F(v_a)&=&\frac{2{\dot Q}({\bf r}_0(v_a))}{\Omega_L}
\left|\tan\frac{\Omega_L t_0}{2}\right|
\nonumber \\
&=&
\frac{{\dot Q}({\bf r}_0(v_a))}{\Omega_L}
\frac{v_a}{v_{w,\perp}}
\left[1-\left(\frac{v_a}{2v_{w,\perp}}\right)^2\right]^{-1/2}, 
\label{eq:flux_cold}
\ea
where ${\bf r}_0$ is related to $v_a$ via equations 
(\ref{eq:r0_t})-(\ref{eq:v_abs}) or (\ref{eq:r_0_2}).
This velocity distribution exhibits certain 
characteristic features. First, it 
goes to zero for particles with the lowest possible velocity 
$v_a\to 0$. Second, $F(v_a)$ exhibits a sharp spike for 
particles with the highest possible velocity $v_a\to 2v_{w,\perp}$. 

In between these velocity extremes, the shape of the $F(v_a)$ 
dependence is modulated by the dependence of the rate 
of particle production in the Oort Cloud $\dot Q$ on the birth distance 
${\bf r}_0$. This property can be used to explore the 
spatial distribution of ${\dot Q}(r_0)$ inside the Cloud,
assuming particle production to be spherically distributed.
Indeed, from equation (\ref{eq:flux_cold}) one derives
\ba
\dot Q(r_0)=F(v_a(r_0))\Omega_L\frac{v_{w,\perp}}{v_a(r_0)}
\left[1-\left(\frac{v_a(r_0)}{2v_{w,\perp}}\right)^2\right]^{1/2},
\label{eq:Qr}
\ea
where the dependence of $v_a$ upon $r_0$ can be extracted
from equation (\ref{eq:r_0_2}). This procedure allows one to 
{\it directly measure the ${\dot Q}(r_0)$ profile} given the 
measurement of the approach velocity distribution $F(v_a)$ in the inner
Solar System. It also informs us of
{\it the direction of the average magnetic field ${\bf B}$ with
respect to the stellar wind velocity ${\bf v}_w$}, because
the upper cutoff of the particle velocity distribution is
directly related to $v_{w,\perp}$.

In performing this exercise one has to remember that at low 
$v_a$, particle motion starts being affected by Solar gravity. 
As a result of gravitational focusing, particle flux measured in 
the ISS is going to be higher than $F(v_a)$ 
(which neglects Solar gravity). To account for this bias, 
we will assume that satellite detectors can measure the
flux $F^{\rm ISS}(v_a,r)$ of particles with {\it approach} 
velocities\footnote{Note that these are not the velocities with which particles strike the detector. Approach velocities $v_a$ can be calculated from energy conservation once the detector position ${\bf r}$ and 
the particle velocities at encounters with spacecraft are known.} in the interval 
$(v_a,v_a+dv_a)$ at a distance $r$ 
from the Sun. These particles arrive on
trajectories with the impact parameter $b$ far from the Sun given by
\ba
b(r)=r\sqrt{1+\frac{2GM_\odot}{r v_a^2}}.
\label{eq:b}
\ea
Since $F^{\rm ISS}(v_a,r)rdr=F(v_a)bdb$, one finds
\ba
F^{\rm ISS}(v_a,r)=F(v_a)\left(1+\frac{GM_\odot}{r v_a^2}\right),
\label{eq:F_ISS}
\ea
which allows one to establish a relation between $F^{\rm ISS}(v_a,r)$
and $\dot Q(r_0)$ via equation (\ref{eq:Qr}). Obviously, the 
focusing correction is small if $v_a^2\gg GM_\odot/r$.


\subsection{Particle arrival direction.}
\label{sect:directions}

Apart from $F(v_a)$, the distribution of particle arrival directions 
also carries important information. We define the orientation of 
the arrival velocity ${\bf v}_a$ near the Sun using two angles: $\eta$ is the 
angle between ${\bf v}_a$ and ${\bf v}_w$ and $\kappa$ is the angle 
between the components of ${\bf B}$ and ${\bf v}_a$ normal to  
${\bf v}_w$. With these definitions it is easy to show that  
\ba
\cos\eta=\frac{1}{v_a}\left({\bf v}_a\cdot {\bf n}_w\right),~~~~~
{\bf n}_w=\frac{{\bf v}_w}{v_w},
\label{eq:theta}
\\
\cos\kappa=\frac{\left({\bf v}_a\cdot {\bf b}\right)-
\left({\bf v}_a\cdot {\bf n}_w\right)
\left({\bf b}\cdot {\bf n}_w\right)}
{\sqrt{v_a^2-\left({\bf v}_a\cdot {\bf n}_w\right)^2}
\sqrt{1-\left({\bf b}\cdot {\bf n}_w\right)^2}}
\label{eq:kappa}
\ea
In particular, in the cold limit we find, using equations 
(\ref{eq:v_t}) and (\ref{eq:v_cold}),
\ba
\cos\eta &=&\sin\frac{\Omega_L t_0}{2} \sin\theta_{wB}=\frac{v_a}
{2v_{w,\perp}}\sin\theta_{wB}=\frac{v_a}{2v_w},
\label{eq:theta1}
\\
\cos\kappa &=& -\frac{\cos\theta_{wB}\sin\left(\Omega_L t_0/2\right)}
{\sqrt{1-\sin^2\theta_{wB}\sin^2\left(\Omega_L t_0/2\right)}}
\nonumber \\
&=&
-\frac{v_a}{2v_w}\frac{\cos\theta_{wB}}{\sin\theta_{wB}}
\left[1-\left(\frac{v_a}{2v_w}\right)^2\right]^{-1/2}.
\label{eq:kappa1}
\ea
These expressions show that the distribution of particle arrival
directions provides us with additional information on the ISM 
magnetic field direction via the dependence on angle $\theta_{wB}$.


\section{Numerical verification of dynamical results}
\label{numerical}


We numerically verified the theoretical predictions obtained in the previous section regarding the dynamics of dust particles originating in the Oort Cloud. To that effect, we integrated trajectories of particles moving in the Solar potential including the electric, magnetic, and Coulomb drag forces, described in \S \ref{perturbations}. Gravitational interactions with planets were not considered, and the Sun was approximated to reside at the center of the potential. Other forces that become important very near the Sun, such as radiation pressure, were also not considered in the simulation. The interstellar wind was approximated as being uniform everywhere, including inside the heliopause. Since we simulated particles for only a few dynamical times, and the interactions with the interstellar wind strongly affect their trajectories only over distances of thousands of AUs (for the particle sizes we consider), these are reasonable approximations. The details of the numerical procedure and simulation setup are described in Appendix \ref{description}.

We randomly populate a cylindrical region of the Oort Cloud (see Appendix \ref{description} for details) between $r_{in}=10^3$ (or $3\times 10^3$ AU) and $r_{out}=10^5$ AU according to the particle production rate $\dot Q(r_0) \propto r_0^{-4}$. This scaling is chosen for verification purposes only --- it is much easier to provide comparison with theory for relatively shallow $\dot Q(r_0)$ profiles. In \S \ref{sect:dust_prod} we explore more realistic profile of $\dot Q(r_0)$.

We run a number of models, each with $N_0=10^{10}$ particles. Given our ``slab'' approach to populating the Cloud described in Appendix \ref{description} this number does not reflect the total number of grains in a spherical Cloud. The latter can be obtained from $N_0$ using equation (\ref{eq:conversion}). At birth, we give the particles a Gaussian initial velocity distribution with dispersion $\sigma_v = 0.7 v_K$ ($v_K(r_0)$ is given by Eq. \ref{eq:v_K}), cut off at $v_{max} = v_{esc}$, and random orientation of the velocity vector. The particle size distribution is $dN/dr_g \propto r_g^{-\zeta}$, and we normally consider only particles in the range between $1$ and $3$ ${\rm \mu}$m in radius. The grain bulk density is $1$ g cm$^{-3}$, and the grain potential is $U=1$ V.

\begin{figure*}[htp]			
\begin{center}
\includegraphics[width=\textwidth]{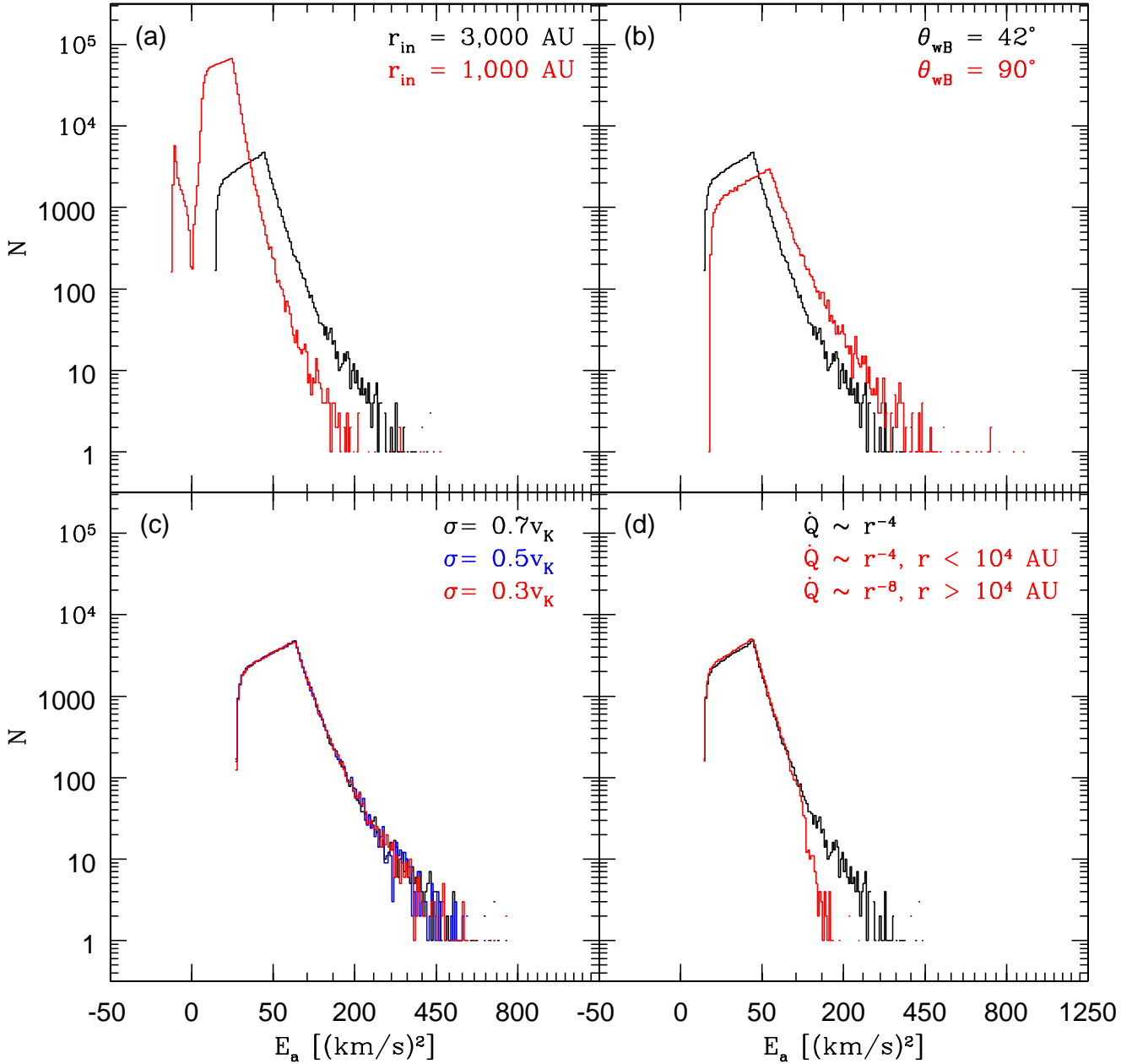} 
\end{center}
\caption{Histograms of specific energy at approach for simulated particles passing within 3-10 AU of the Sun. Orbits of $10^{10}$ particles were integrated. The black curve is the same in all panels and corresponds to $r_{in}=3,000$ AU, $\theta_{wB} = 42^\circ$, $\sigma_v = 0.7v_K$ and $\dot Q\propto r_0^{-4}$. 
Panel (a): comparison of simulations with different inner edges to the Oort Cloud of 1,000 and 3,000 AU. Bound particles appear in the case of $r_{in} = 1,000$ AU with negative approach energies and will generally return to the inner Solar System.
Panel (b): comparison of simulations with $\theta_{wB} = 42^\circ$ and $\theta_{wB} = 90^\circ$. In both cases, the maximum approach velocity is $\approx 2v_{w,\perp}$. 
Panel (c): comparison of simulations with different initial velocity dispersions of $\sigma_v$ = 0.7, 0.5, and 0.3 $v_K$. 
Panel (d): comparison of an $r_0^{-4}$ dust production profile with the one given by equation (\ref{eq:broken_PL}). The histogram correctly recovers the much lower resultant flux of high-velocity particles originating from the outer parts of the Cloud.} 
\label{E4}
\end{figure*}

We set the parameters of the ISM to the measured wind speed of 23.2 km s$^{-1}$ \citep{McComas} and magnetic field strength of 2.2 ${\rm \mu}$G \citep{Ben-Jaffel}, in accordance with our adopted values in Table \ref{ISMmodels}. The field orientation is different for different models, but in most cases, we use the measured value $\theta_{wB} = 42^\circ$ from \citet{Ben-Jaffel}. In ecliptic coordinates, the wind direction lies at (255.4$^\circ$, 5.2$^\circ$), and the magnetic field is at (224$^\circ$, 36$^\circ$) \citep{Ben-Jaffel}. For the other relevant parameters, we set the ISM density \citep{McComas,Ben-Jaffel} to $n = 0.07$ cm$^{-3}$, the temperature to $T = 6300$ K; we use the approximate value ${\cal F} = 20$ for the warm ISM phase (BR10). 

To achieve better statistics in our simulations we count the number of particles $N(v_a,r_1<r<r_2)$, with approach velocities in the interval $(v_a,v_a+dv_a)$ (although we plot them in terms of approach energy, see below) having closest approaches to the Sun between $r_1$ and $r_2$ per unit time $dt$. Using the relation (\ref{eq:b}) between the closest approach distance $r$ and the impact parameter $b$ far from the Sun, one can express
\ba
N(v_a,r_1<r<r_2)=\pi\left[b^2(r_2)-b^2(r_1)\right]F(v_a)dt.
\label{eq:N}
\ea
This relation allows us to express the particle approach velocity distribution $F(v_a)$ via the measured $N(v_a,r_1<r<r_2)$, which can then be plugged into equation (\ref{eq:Qr}). In the limit $dr=r_2-r_1\ll r_{1,2}\approx r$, one finds $N(v_a,r_1<r<r_2)\approx 2\pi r F^{\rm ISS}(v_a,r)dr$, and expression (\ref{eq:F_ISS}) is easily recovered from equation (\ref{eq:N}).

In checking the analytical predictions of \S \ref{sect:small_dyn},
we pay special attention to the sensitivity of our numerical 
results to various parameters of the model, such as the 
magnetic field direction, parameterized by the angle 
$\theta_{wB}$, the location of the inner edge of the Cloud 
$r_{in}$, and the form of the particle production function 
$\dot Q(r_0)$.

\begin{figure}[htp]			
\begin{center}
\includegraphics[bb = 18 400 330 718,clip,width=\columnwidth]{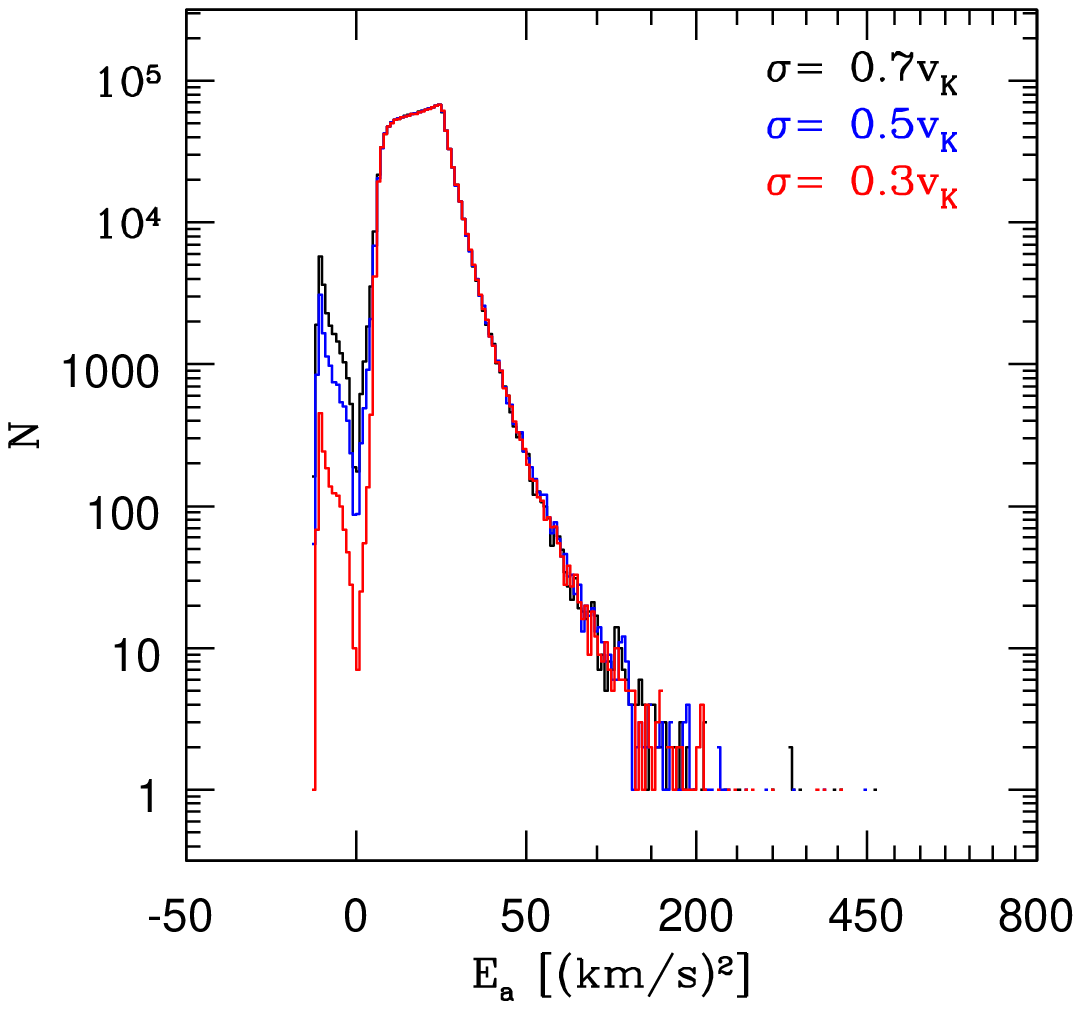}
\end{center}
\caption{Same as Figure \ref{E4}c but for $r_{in} = 10^3$ AU, again with different starting velocity dispersions of $\sigma_v$ = 0.7, 0.5, and 0.3 $v_K$. The number of bound particles (peaking at negative energies) rapidly increases with $\sigma_v$, see the text for details.}
\label{Esigma}
\end{figure}


\subsection{Particle velocity distribution in the ISS}
\label{vel_dist}

We start by presenting distributions of particle approach velocities $v_a$ in the ISS as they would be measured by satellite dust detectors, i.e. by fully taking into account the effect of gravitational focusing of low-velocity particles. Since some of the particles are bound, we express the velocity distribution via their specific energy at approach (rather than their approach velocity $v_a$), which is negative for bound particles and positive for unbound grains ($E_a = (1/2)v_a^2$ for the latter), and is measured in (km/s)$^2$. In Figure \ref{E4}, we plot histograms of $N(E_a,r_1<r<r_2)$ (linearly in $v_a$) given by equation (\ref{eq:N}) for $r_1=3$ AU and $r_2=10$ AU and different model parameters.

In panel (a) of Figure \ref{E4}, we explore the effect of changing the inner radius of the Oort Cloud on the distribution of particle energies. Plotted are histograms of the specific energy at approach, for two simulations: one with $r_{in}=1,000$ AU and one with $r_{in}=3,000$ AU. At high energies, which correspond to larger approach velocities and thus larger birth distances, particle numbers drop rapidly. Most particles have low velocities; the considerable width of the peak in their numbers is due to the finite range of particle sizes used in the simulations (see equation \ref{eq:v_min}). The velocity peak is shifted to lower energies in the $10^3$ AU case because most particles are accelerated over shorter distances compared with the $3,000$ AU case. Also, some particles end up bound in the 1,000 AU case, appearing with negative energies as measured in the inner Solar System. Their subsequent fate may depend on a number of factors. Some of them have large semi-major axes $\sim 10^3$ AU and are likely to be accelerated by the ISM induction drag ultimately removing them from the Solar System. Other, more strongly bound particles, would return to the inner Solar System after one orbital time, and may be affected both by planetary perturbations and by the Solar wind. Their fate is hard to predict. However, we are primarily interested in unbound particles, for which this issue does not apply.

The fraction of grains produced inside the cloud that pass within 3-10 AU from the Sun is $\approx 9\times 10^{-5}$ and $\approx 3\times 10^{-6}$ in the $r_{in}=1,000$ AU and $r_{in}=3,000$ AU cases, respectively (applying the ``slab'' correction given by equation (\ref{eq:conversion})). This difference is in rough agreement with the scaling of the particle flux $\propto r_{in}^{-3}$, which is not too far from the theoretical prediction (\ref{eq:FISS_full_est}).

In panel (b), we compare the effect of a magnetic field orientation of $\theta_{wB} = 90^\circ$ with our adopted value of $\theta_{wB} = 42^\circ$. This introduces obvious differences --- one can see particles moving with speeds up to $46$ km s$^{-1}$ ($E = 1060$ km$^2$ s$^{-2}$) in the former case, while in the latter case particle velocities do not exceed $31$ km s$^{-1}$ ($E_a = 480$ km$^2$ s$^{-2}$). This is in agreement with analytical predictions (\ref{eq:flux_cold}), according to which the maximum approach velocity is $2v_{w,\perp}=2v_{w,\perp}\sin\theta_{wB}$. This is $\approx 46$ km s$^{-1}$ for $\theta_{wB}=90^\circ$ and $\approx 31$ km s$^{-1}$ for $\theta_{wB}=42^\circ$. The peak in the approach velocity distribution similarly occurs at larger velocities for $\theta_{wB} = 90^\circ$.

In panel (c), we compare simulations with different values of the initial velocity dispersion parameter, $\sigma_v$, specifically, $\sigma_v = 0.7v_K, 0.5v_K,$ and $0.3v_K$. This has little effect on the unbound particles occurring with $r_{in} = 3,000$ AU, which are quickly accelerated to high speeds and swept out of the Solar System. However, we find that in the $r_{in} = 1,000$ AU case, the number of bound particles is significant and increases rapidly with the velocity dispersion, see Figure \ref{Esigma}. We observe $\approx 2\times 10^{4}$ bound (1.7\% of the total) particles passing at 3-10 AU for $\sigma_v = 0.7 v_K$. This reduces to $\approx 1.1\times 10^4$ (0.84\% of the total) and $1.6\times 10^3$ (0.12\% of the total) for $\sigma_v = 0.5 v_K$ and $0.3 v_K$, respectively.

In panel (d), we test the sensitivity of $N(v_a,r_1<r<r_2)$ to features in the radial dependence of the particle production rate $\dot Q$. We design a simulation with $\theta_{wB}=42^\circ$ and $r_{in}=3,000$ AU in which
\ba
   \dot Q(r_0) \propto \left\{
     \begin{array}{lr}
       r_0^{-4}, ~~~~3\times 10^3 {\rm AU} < r_0 < 10^4 {\rm AU}\\
       r_0^{-8}, ~~~~10^4 {\rm AU} < r_0 < 10^5 {\rm AU}
     \end{array}
   \right.
\label{eq:broken_PL}
\ea
and compare it with a simulation in which $\dot Q(r_0) \propto r_0^{-4}$ for $3,000 {\rm AU} < r_0 < 10^5 {\rm AU}$. One can see that for low $E_a\lesssim 10^2$ (km/s)$^2$, the two energy distributions are nearly identical, which is consistent with the fact that low-energy particles are produced in the inner part of the Cloud. Higher energy grains are created beyond $10^4$ AU, where the broken power law production rate is steeply decaying with $r_0$, explaining a very different behavior of that particle energy distribution --- $N(v_a,r_1<r<r_2)$ falls much faster with $E_a$ at high $E_a\gtrsim 100$ (km/s)$^2$ than for $\dot Q \propto r_0^{-4}$.

There are several generic features of $N(v_a,r_1<r<r_2)$ that persist through all simulations shown in  Figure \ref{E4}. First, all simulations show that most particles have low $v_a\lesssim 10$ km s$^{-1}$, with $N(v_a,r_1<r<r_2)$ decaying as $v_a$ increases. This is because higher $v_a$ means larger grain birth distance $r_0$, implying a lower production rate $\dot Q$. Most of the low-velocity particles are produced close to the inner edge of the Cloud. Their velocity distribution is sensitive to the value of $r_{in}$. While the case of $r_{in} = 3,000$ AU gives most of the particles $v_a \lesssim 10$ km s$^{-1}$, $r_{in} = 1,000$ AU and places most of the particles at $v_a \lesssim 5$ km s$^{-1}$.

Second, all our runs with $r_{in} = 1,000$ AU exhibit a considerable number of bound particles showing up as the spike at negative $E_a$, see Figure \ref{Esigma}. These particles have apparently originated at $r < 3,000$ AU and have not been accelerated by the ISM wind enough to completely unbind them from the stronger Solar gravity in that region. The variation in $\sigma_v$ dramatically changes the number of these bound particles.

Overall, we find that the numerical results for $N(v_a,r_1<r<r_2)$ presented here are in good qualitative agreement with the analytical expectations in \S \ref{sect:small_dyn}, despite the use of the zero initial velocity assumption.


\subsection{Particle arrival directions}
\label{vel_dir}

Next, we compare the arrival directions of the simulated particles with the analytical predictions from Equations (\ref{eq:theta1}) and (\ref{eq:kappa1}). Again, it has to be remembered that we are discussing the direction of the {\it approach} velocity in the ISS, rather than the grain speed measured by the satellite, which is affected by gravitational focusing (knowledge of the latter uniquely determines $v_a$).

\begin{figure*}[htp]			
\begin{center}
\subfigure{\includegraphics[bb = 14 50 947 602,clip,width=0.49\textwidth]{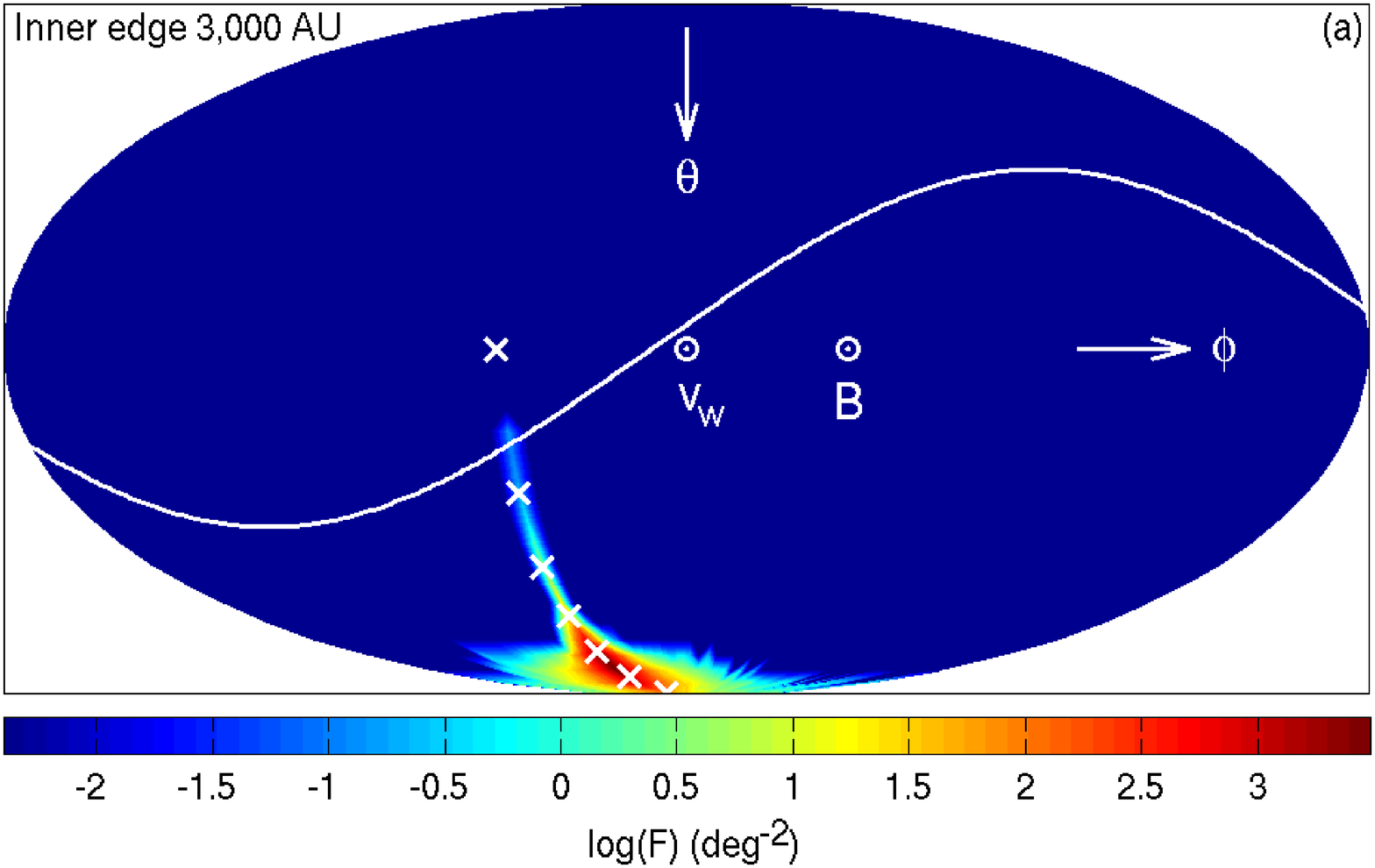}}
\subfigure{\includegraphics[bb = 14 50 947 602,clip,width=0.49\textwidth]{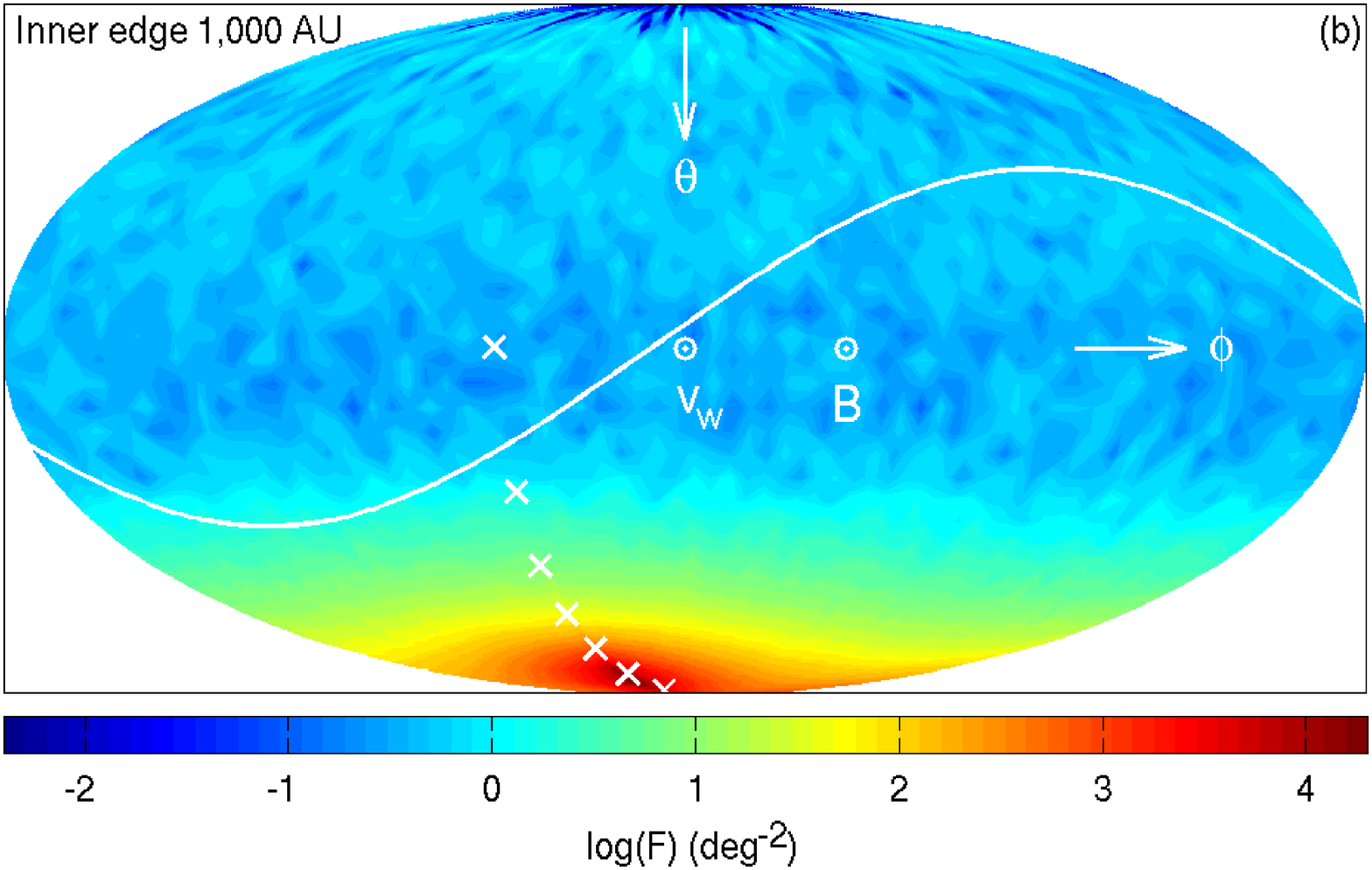}}
\subfigure{\includegraphics[bb = 14 14 828 527,clip,width=0.49\textwidth]{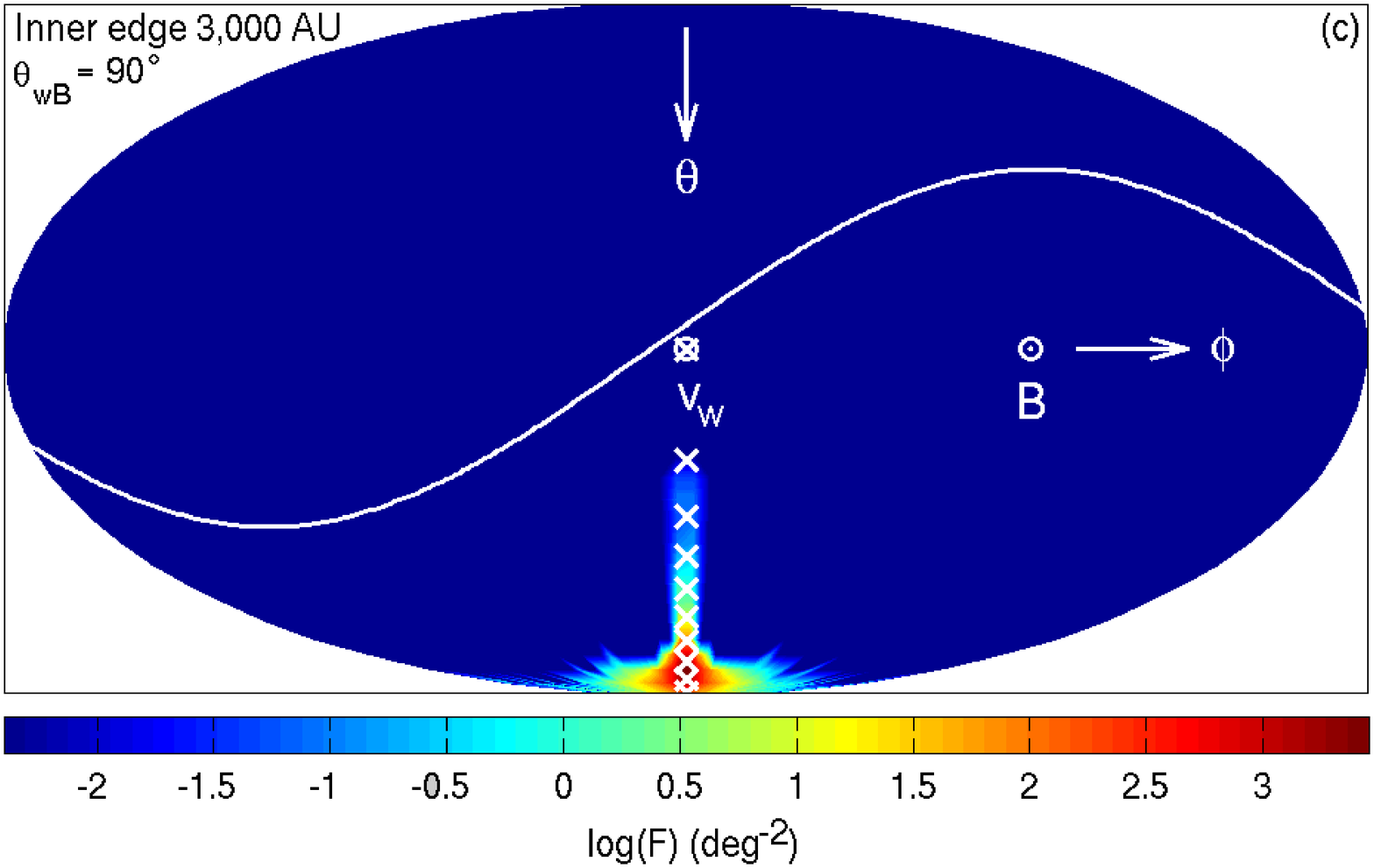}}
\subfigure{\includegraphics[bb = 14 14 828 527,clip,width=0.49\textwidth]{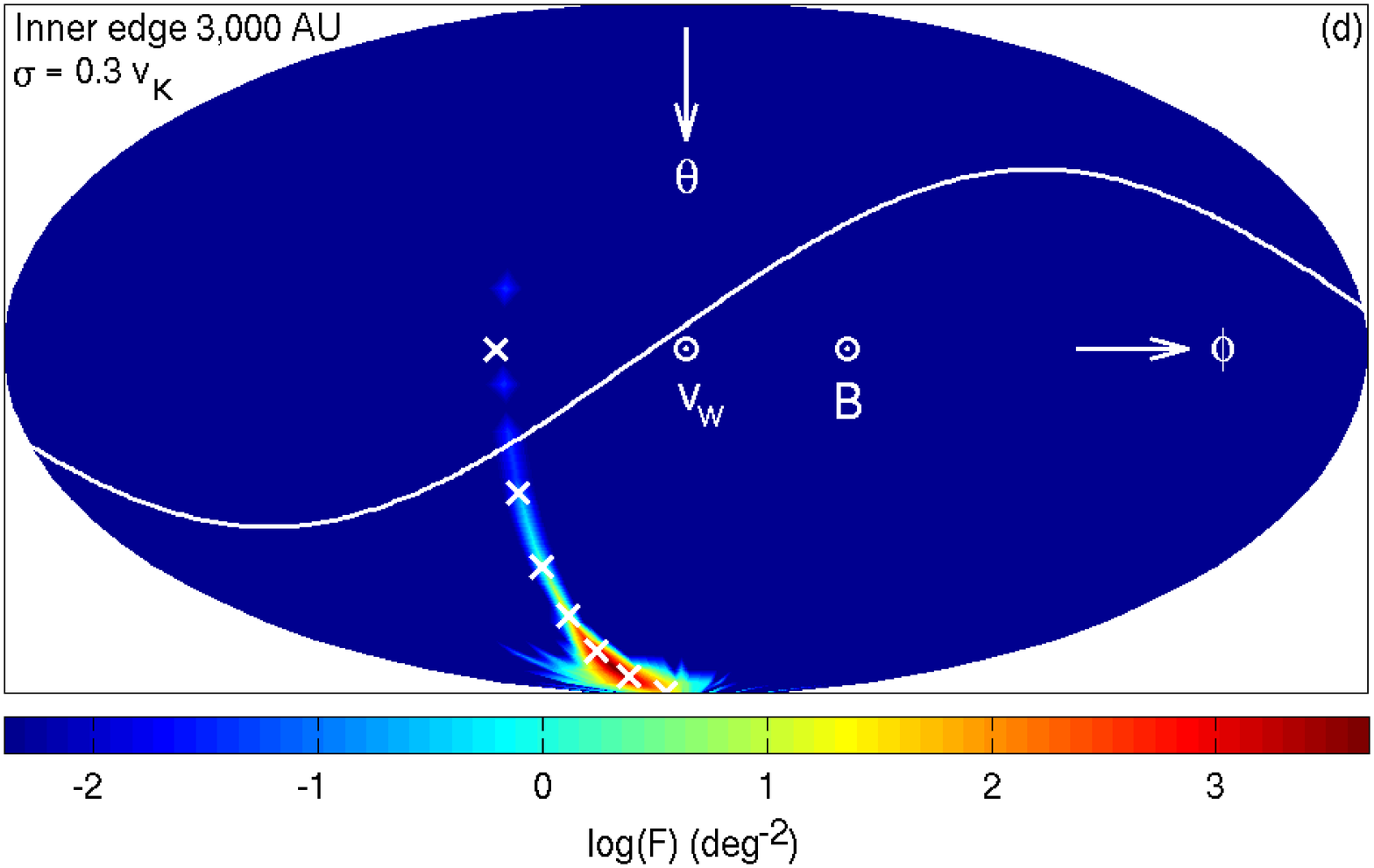}}
\end{center}
\caption{Asymptotic arrival directions of simulated particles approaching within 3-10 AU of the Sun in $\theta-\phi$ coordinates described in the text. In these and all subsequent maps, the interstellar wind arrives from $\theta = 90^\circ$ and $\phi = 0^\circ$; the ISM magnetic field direction is in the $\theta = 90^\circ$-plane. The resolution of arrival directions is 5$^\circ$. Color corresponds to the logarithm of the flux in particles per square degree ($10^{10}$ orbits were integrated in total). In all cases, cometary particles arrive predominantly from the pole at $\theta=180^\circ$, i.e. perpendicular to both ${\bf B}$ and ${\bf v}_w$. Panels (a) and (b): comparison of simulations with $r_{in}=3,000$ and 1,000 AU. In the 1,000 AU case, bound particles contribute a roughly isotropic background to the map, see Figure \ref{thvolts}b. Panel (c): simulation with $\theta = 90^\circ$, correctly recovering the perpendicular distribution of arrival directions. Panel (d): simulation with smaller velocity dispersion of $\sigma_v$ = 0.3 $v_K$, which somewhat reduces the scatter of arrival directions.}
\label{map4}
\end{figure*}

We use a polar coordinate system ($\phi$,$\theta$) on the sky in which the ISM wind direction lies at $\phi = 0^\circ$ and $\theta = 90^\circ$ and the magnetic field direction lies in the $\theta = 90^\circ$ plane. In Figure \ref{map4}, we plot the direction of ${\bf v}_a$ in these coordinates in a Mollweide projection. The white curve marks the ecliptic plane, and the colors correspond to $\log$(number density of particles). The angular resolution of these plots is $5^\circ$. The crosses on each plot mark the analytic arrival directions for velocity bins with widths of 5 km s$^{-1}$, and the wind velocity and magnetic field directions are marked. The ecliptic is shown by a white curve.

The maps show that all of the particles, bound and unbound, arrive roughly perpendicular to the magnetic field direction when $r_{in} = 3,000$ AU. This is to be expected because of the induction force, which causes the motion of particles to be perpendicular to the magnetic field direction. For $\theta_{wB} = 90^\circ$, the perpendicular directions correspond to the vertical midline (and the $\phi = 0^\circ$ half-plane), while for $\theta_{wB} = 42^\circ$, the plane of particle motion appears on the plot as an ovoid circumscribed around that point, although particles only arrive along an arc of that ovoid. In these simulations, the particle arrival directions fall very near the expected analytic distribution.

When $r_{in} = 1,000$ AU, a very different behavior occurs. The deeper potential well at the inner edge of the Cloud results in many of the particles remaining bound, or only marginally unbound. Their larger initial velocities (because of higher $v_K$) make the cold limit less accurate. As a result of these two factors, the expected angular distribution is somewhat washed out, and particles arrive from a wide region of the sky centered on the normal to the $v_w-B$-plane, characteristic of low approach energies, as expected given the small birth distance for most of the particles. A significant number of particles arrive from the opposite direction on the sky; these turn out to be bound particles only (see below). In some of the maps a small overdensity of particles in the plane perpendicular to the magnetic field direction remains visible; these are the few particles that are accelerated to high velocities and swept along in the direction of the interstellar wind.

\begin{figure}[htp]			
\begin{center}
\includegraphics[bb = 18 144 330 718,clip,width=0.45\textwidth]{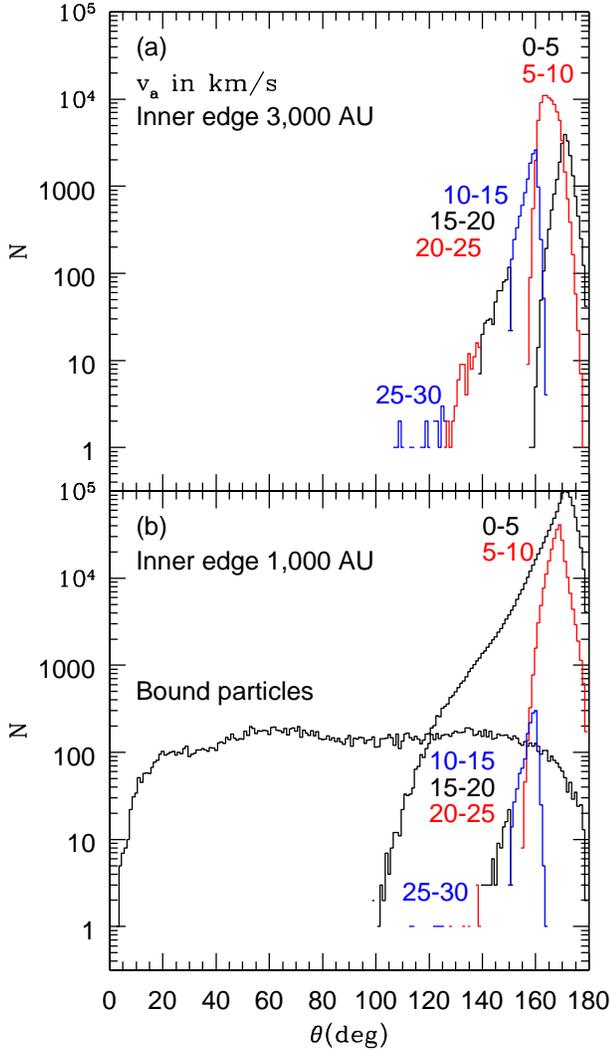}
\end{center}
\caption{Histograms of particle arrival directions in the $\theta$-coordinate integrated over $\phi$, broken down into 5 km s$^{^-1}$ wide approach velocity bins, as indicated in the panels. Results from simulations with (a) $r_{in}=3,000$ AU and (b) $r_{in}=1,000$ AU are shown with 1$^\circ$ resolution. The 1,000 AU case results in a significant number of bound particles with a roughly uniform directional distribution.}
\label{thvolts}
\end{figure}

For clarity and additional analysis, we show the projections of these plots onto the $\phi = 0^\circ$ half-plane in Figure \ref{thvolts} with the addition of breaking them down by approach velocity bins. Taking advantage of better statistics, we use an angular resolution of $1^\circ$ for these plots to illustrate arrival directions of particles of different speeds.

These projected direction plots illustrate that the arrival directions of bound particles are roughly uniform across the sky. Marginally unbound particles with low positive approach energies (low real approach velocities) also have a relatively wide distribution of arrival directions.


\subsection{Reconstruction of $\dot Q$ behavior}
\label{Qdot}

\begin{figure*}[htp]			
\begin{center}
\includegraphics[width=\textwidth]{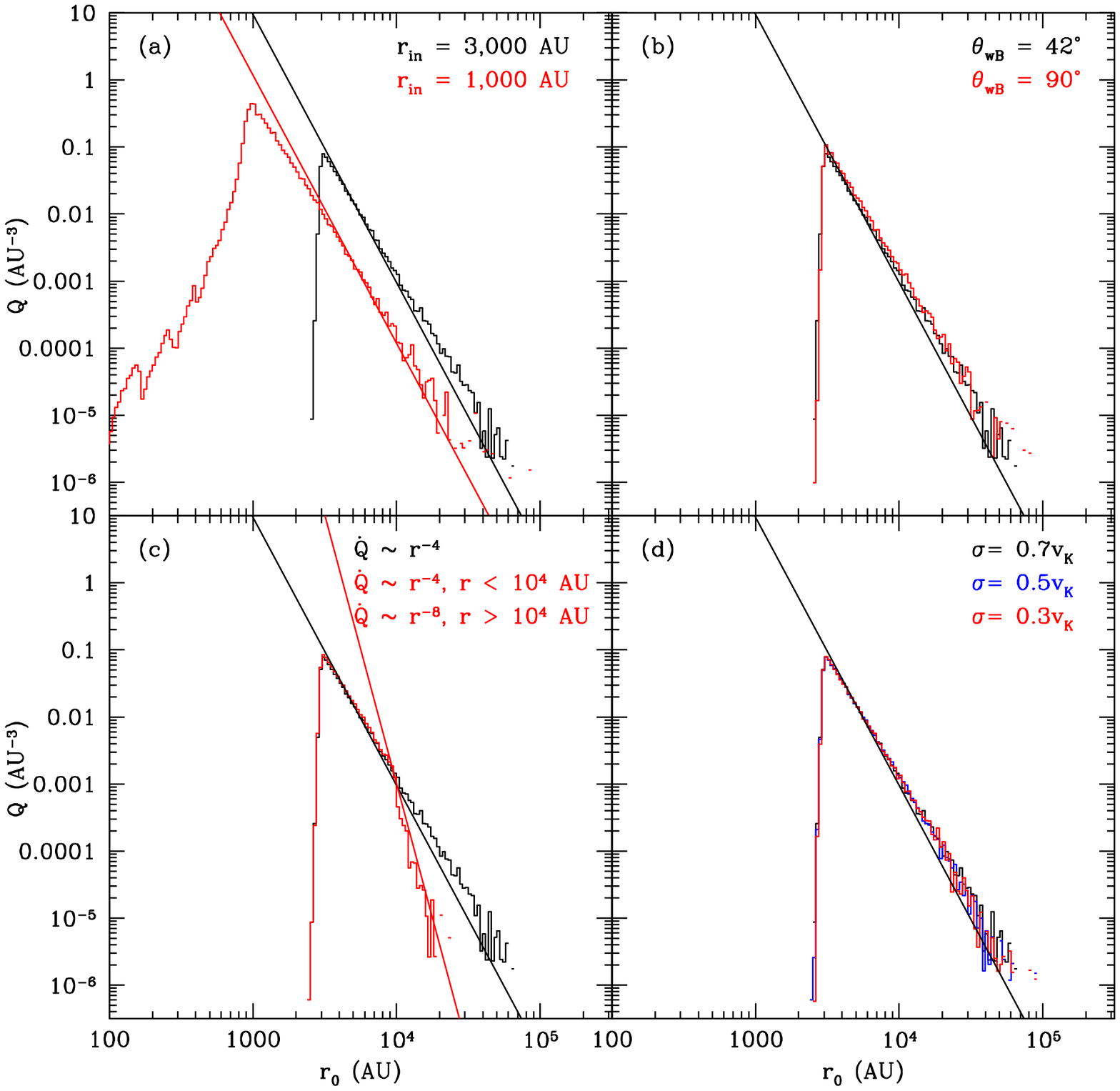} 
\end{center}
\caption{Reconstruction of particle creation profiles (histograms) $Q(r_0)$ (as a proxy for $\dot Q$) based on velocity distributions of simulated particles ``observed'' near the Sun. Diagonal lines represent the slopes of the input $Q(r_0)$ used in simulations. The black curve is identical in each case. Panel (a): comparison of simulations with $r_{in}=3,000$ (fiducial choice) and 1,000 AU. In the 1,000 AU case, bound particles affect reconstruction near the inner edge. (b) Comparison for different magnetic field directions of $\theta_{wB} = 42^\circ$ and $90^\circ$. (c) Comparison of $\dot Q\propto r_0^{-4}$ with the one given by equation (\ref{eq:broken_PL}). The reconstruction successfully recovers the feature in the radial distribution of particles--the broken power law. (d) Reconstruction for different initial velocity dispersions $\sigma_v = 0.7, 0.5,$ and $0.3 v_K$. In all cases, the peak of the reconstructed profile occurs at the modeled inner edge of the Oort Cloud.}
\label{r4}
\end{figure*}

Finally, based on our simulations, we illustrate how the measured particle velocity distributions can be used to infer basic properties of the Oort Cloud. Namely, we use equation (\ref{eq:Qr}) to reconstruct the radial profile of the dust production rate $\dot Q(r_0)$ based on the measurement of $N(v_a,r_1<r<r_2)$ in our simulations, see \S \ref{vel_dist}. The latter is used to obtain $F(v_a)$ via equation (\ref{eq:N}), which then allows us to infer $\dot Q(r_0)$ using equation (\ref{eq:Qr}). The results of applying this procedure to the cases shown in Figures \ref{E4} are demonstrated in Figures \ref{r4}, using $r_1 = 3$ AU and $r_2 = 10$ AU, as before.

The thin lines indicate the slope of the actual behavior of $Q(r_0)$ used as an input in simulations. One can see that the reconstructed $Q(r_0)$ is in very good agreement with the input behavior, which validates our procedure. Not only are we able to reproduce the slope of $Q(r_0)$ quite well, irrespective of the value of $\theta_{wB}$, but we can also infer the existence of the break in the behavior of $\dot Q(r_0)$ in the cases shown in panels (b) and (c) Figure \ref{r4}.

One slight issue with our reconstruction procedure is that it yields particle production inward of the real input rim. This is very obvious for $r_{in} = 1,000$ AU, but it also occurs for $r_{in} = 3,000$ AU, as shown in panel (a) of Figure \ref{r4}. We suspect this to be due to the initial velocity of particles at birth $v_0$ not being negligible compared with the velocity gain due to ISM wind coupling. Indeed, this problem arises mainly at small $r_0$, which corresponds to particle approach velocities of $\lesssim 10$ km s$^{-1}$ and is more pronounced for small $r_{in}$. For these grains, the non-zero initial velocity of order the local Keplerian velocity makes a difference. Neglecting it by adopting the cold limit as we do in our reconstruction procedure is likely to affect the reconstructed value of $r_{in}$.

Nevertheless, we generally find that the measurement of the particle velocity distribution in the ISS allows us to reconstruct the features of the radial dependence of $\dot Q(r_0)$ as long as the statistics are sufficient.


\section{Dust production in the Oort Cloud}
\label{sect:dust_prod}


We now explore dust production in collisions between the Oort Cloud
comets and make predictions for the dust flux that may
be measured by space-borne missions. 

We assume that comets are spherically-symmetrically distributed in
the Cloud with initial mass density $\rho_0(r_0)\propto r_0^{-\eta}$,
where $\eta\approx 3.5-4$ as suggested by numerical calculations 
of the Oort Cloud's formation \citep{Duncan}. Analogous to
\S \ref{sect:small_dyn}, we use $r_0$ for the distance in the 
Oort Cloud. Such a Cloud contains 
most of its mass at the inner edge, located at radius $r_{in}$. 
For a given total mass of the Cloud, $M_{\rm OC}$, we then find
\ba
\rho_0(r_0)\approx \frac{\eta-3}{4\pi}\frac{M_{\rm OC}}{r_{in}^3}
\left(\frac{r_{in}}{r_0}\right)^\eta.
\label{eq:rho_0}
\ea
Assuming that the initial population has most of its mass in objects
of size $d_m$, one finds the initial number density of comets to be
\ba
n_0(r_0)\approx 10^3\mbox{AU}^{-3}(\eta-3)\frac{M_{\rm OC,10}}{d_{m,1}^3
r_{in,3}^3}\left(\frac{r_{in}}{r_0}\right)^\eta,
\label{eq:n_0}
\ea
where $M_{\rm OC,10}\equiv M_{\rm OC}/(10M_\oplus)$, 
$d_{m,1}\equiv d_m/1$ km, and $r_{in,3}\equiv r_{in}/10^3$ AU.

For simplicity, we will assume that all comets move on circular 
orbits. This eliminates their eccentricity from the problem and
means that at a given radius, the process of fragmentation is not affected by
other parts of the Cloud, i.e. the motion of all comets is confined to
spherical shells. The rms collision speed in this case is 
$v_r=v_K/\sqrt{2}$, where $v_K$ is the Keplerian speed.


\subsection{Collisional evolution of the Oort Cloud}
\label{subsect:coll}

As a result of collisions between comets, their mass spectrum 
$dn/dm$ --- the number density of comets in the mass interval 
$(m,m+dm)$ --- develops a fragmentation tail at low masses, 
which steadily extends to larger and larger values of $m$. 
At any given time the largest mass that is a part
of the fragmentation cascade is the mass $m_{coll}(t)$ for which the 
collision probability integrated over the lifetime $t$ of the system is 
equal to unity. The mass flux in the collisional cascade per 
unit volume $dF_m(m)/dV$, i.e. the mass of the collisional products crossing 
a given $m$ in mass space per unit time and unit volume, is independent of the mass of the body for $m<m_{coll}(t)$. As a
result, mass of collisional debris joining the cascade at 
$m\sim m_{coll}$ simply flows toward smaller and smaller 
$m$ where some physical mechanism usually provides a mass sink 
(see below). This statement implicitly assumes that no mass mixing  
between different physical volumes of space occurs, i.e. that debris
produced at some radius stays there. 

As time goes by, $m_{coll}$ reaches $m_m=(4\pi/3)\rho d_m^3$ 
--- the mass scale containing most of the mass of the comet 
population ($\rho$ is the bulk density of comets, which
we assume to be equal to 1 g cm$^{-3}$). This happens at 
time equal to the {\it erosion timescale} $t_{er}$ defined via 
$m_{coll}(t_{er})= m_m$. Prior to $t_{er}$ the total mass 
of the comet population stays roughly the same, while $m_{coll}$
increases. Later on, for $t\gtrsim t_{er}$, the local density in comets  
decays with time, while $m_{coll}\approx m_m$. 
This process can be described using a simple evolution equation 
for the local mass density $\rho(r,t)$ of the comet population:
\ba
\frac{d\rho}{dt}\approx -\rho\times \langle\sigma v\rangle
\frac{\rho}{m_m}\approx -\frac{\rho^2}{\rho_0 t_{er,0}},
\label{eq:ev} 
\ea
where we have introduced the erosion time at the beginning of Cloud evolution 
$t_{er,0}^{-1}\approx n_0\times 
\langle\sigma v\rangle\approx \rho_0\times \langle\sigma v\rangle/m_m$.
Solving equation (\ref{eq:ev}), we find the following behavior of 
the local mass density of comets with time:
\ba
\rho(r_0,t)\approx \rho_0(r_0)\left[1+\frac{t}{t_{er,0}(r_0)}\right]^{-1}.
\label{eq:mev}
\ea
Analogous results were obtained by \citet{Dominik} and \citet{Wyatt07}.

The erosion timescale is naturally a function of the distance $r_0$ 
from the Sun, among other things. Here we assume that $t_{er}$ 
scales with distance as 
\ba
t_{er}(r_0)=t_{er}(r_{in})\left(\frac{r_0}{r_{in}}\right)^{\kappa},
\label{eq:ter_r}
\ea
where $\kappa \approx 4-5$ based on the calculations presented in
Appendix \ref{sect:erosion}, which motivates the dependence (\ref{eq:ter_r}).

The mass flux in the collisional cascade at time $t$ can be 
evaluated as 
\ba
\frac{dF_m}{dV}\approx \frac{\rho(m_{coll}(t))}{t}\approx
t^{-1}\left(m^2\frac{dn}{dm}\right)\Big|_{m=m_{coll}(t)},
\label{eq:mass_flux}
\ea
where $\rho(m)\sim m^2 dn/dm$ is the volume density of objects of mass 
$\sim m$ (in a mass interval of width $\sim m$).
An important characteristic value of the mass flux 
$dF_{m,0}/dV$ is achieved at $t=t_{er}$ when $\rho(m) = \rho_0$:
\ba
\frac{dF_{m,0}(r_0)}{dV}&\approx& \frac{\rho_0(r_0)}{t_{er}(r_0)}
\nonumber \\
&\approx& \frac{\eta-3}{4\pi}\frac{M_{\rm OC}}{r_{in}^3t_{er}(r_{in})}
\left(\frac{r_{in}}{r_0}\right)^{\eta+\kappa},
\label{eq:Fm0}
\ea
where we used equation (\ref{eq:ter_r}). 

We will now assume for simplicity that the initial size distribution 
of comets is the same as the size distribution in fragmentation 
cascade. In that case, the mass flux at low $m$ stays the same and equal to
$dF_{m}(r_0,t)/dV=dF_{m,0}(r_0)/dV$ as long as $t\lesssim t_{er}$. For  
$t\gtrsim t_{er}$ local cometary density starts to decay according
to equation (\ref{eq:mev}), so that $\rho(m_{coll})\approx\rho(m_m)
\approx \rho_0(t/t_{er})^{-1}$, since most of the cometary mass is
in the largest bodies with mass $m_m$. As a result, equation 
(\ref{eq:mass_flux}) predicts that $dF_{m}(r_0,t)/dV=
\left(dF_{m,0}(r_0)/dV\right)(t/t_{er})^{-2}$ at late times.  
These asymptotic behaviors motivate the following 
simple prescription for the mass flux evolution at all times:
\ba
\frac{dF_m(r_0,t)}{dV}\approx \frac{dF_{m,0}(r_0)}{dV}
\left[1+\frac{t}{t_{er}(r_0)}\right]^{-2},
\label{eq:Fmev}
\ea
which we employ in this work. In this approximation, 
the mass flux stays at the high level $dF_{m,0}(r_0)/dV$ as
long as $t\lesssim t_{er}$ but then rapidly decays with time. 

If the initial mass spectrum is different from that of the 
fragmentation cascade, then $dF_m(r_0,t)/dV$ will differ from 
equation (\ref{eq:Fmev}) at 
$t\lesssim t_{er}(r_0)$ (i.e. $dF_{m,0}(r_0)/dV$) by a 
multiplicative factor $(t/t_{er})^\nu$, where $\nu$ 
is a small power --- a detail that we neglect here for 
simplicity.


\subsection{Production of small unbound dust particles}
\label{subsect:smalldust}

Satellite dust detectors can measure properties of only
small, ${\rm \mu m}$-sized particles. Such particles do not belong 
to the fragmentation cascade 
as they are not bound to the Sun (BR10): the induction electric force due to 
the magnetized ISM wind unbinds the particles with sizes
smaller than $r_{g,min}$, estimated in equation (\ref{rgminE}).
This estimate is accurate as long as particles 
are born with relatively low velocities compared with 
the local Keplerian value, $v_0\lesssim 0.5 v_K$, but 
remains reasonably good even for somewhat higher $v_0$.

Equation (\ref{rgminE}) implies that at distance $r_0$ from 
the Sun, the fragmentation cascade cuts off at a minimum 
particle size 
\ba
r_{g,min}(r_0)=r_{g,min}(r_{in})\frac{r_0}{r_{in}}.
\label{eq:d_cut}
\ea
Particles with $r_g< r_{g,min}$ produced in grain-grain collisions 
at the low-mass end of the cascade get accelerated by the 
induction electric force and rapidly removed from the Cloud on a 
dynamical timescale. Their size distribution should then be 
close to that of the size spectrum of fragments produced in 
a {\it single} particle fragmentation event, which is typically 
found to be roughly matched by a power law \citep{Fujiwara}. 
For that reason we assume that unbound fragments have a size 
distribution scaling as $\propto r_g^{-\zeta}$ with $\zeta<4$ 
(so that most of the debris mass produced in a single collision 
is in largest particles) for $r_g< r_{g,min}$. Experimental 
results suggest $\zeta \approx 3.5$ \citep{Fujiwara}.

The production rate of particles of a given size $\dot Q(r_g)$, i.e.
the number of particles of size $r_g$ per unit $r_g$ volume, and time, 
also scales as $r_g^{-\zeta}$. Normalizing the total mass 
in unbound particles created per unit time and unit volume 
$\int_0^{r_{g,min}}\dot Q(r_g)m_g dr_g$ to the mass flux in the 
fragmentation cascade $dF_m(r,t)/dV$, one finds that 
\ba
\dot Q(r_0,t,r_g) &= &\frac{3(4-\zeta)}{4\pi\rho r_{g,min}^4}
\frac{dF_m(r_0,t)}{dV}\left[\frac{r_{g,min}(r_0)}{r_g}\right]^\zeta
\label{eq:dotQ}\\
&=& \frac{3(4-\zeta)(\eta-3)M_{\rm OC}}
{16\pi^2\rho r_{g,min}^4(r_{in})r_{in}^3t_{er}(r_{in})}
\left[\frac{r_{g,min}(r_{in})}{r_g}\right]^\zeta
\nonumber \\
&\times&
\left(\frac{r_{in}}{r_0}\right)^{4+\eta+\kappa-\zeta}
\left[1+\frac{t}{t_{er}(r_0)}\right]^{-2}
\label{eq:dotQ1}
\ea
Note that since $r_{g,min}$ increases with the distance from 
the Sun and $\zeta<4$, the production rate falls off with $r_0$
even more steeply than the mass flux. In parts of the Cloud where 
the local erosion time is longer than the Solar System age $t$ 
(expected to be the case in most if not all of the OSS, see Appendix 
\ref{sect:erosion}), i.e. $t\lesssim t_{er}(r_0)$ one finds 
$\dot Q\propto r_0^{\zeta-(4+\eta+\kappa)}$. In the innermost 
parts of the Cloud, where the opposite condition 
$t\gtrsim t_{er}(r_0)$ may be fulfilled, equation (\ref{eq:dotQ1}) 
predicts a much shallower dependence 
$\dot Q\propto r_0^{\zeta+\kappa-(4+\eta)}$. This results 
in a broken power law behavior of $\dot Q(r_0)$ --- a situation 
illustrated by equation (\ref{eq:broken_PL}) and Figures \ref{E4}d \& \ref{r4}d.


\subsection{Flux of cometary dust particles in the ISS}
\label{sect:flux_ISS}

Equations (\ref{eq:Fm0}) and (\ref{eq:dotQ}) demonstrate that 
$\dot Q$ very rapidly decays with the distance. As a result, we 
expect that most particles that can be detected by satellite 
dust detectors would originate from the innermost part of the 
Cloud, at $r\sim r_{in}$. Because $r_{in}\ll R_L\sim 10^4$ AU for 
$\mu$m-sized fragments, see equation (\ref{eq:R_L}), these 
particles are relatively weakly accelerated by the induction 
electric force and appear in the ISS as a low-velocity 
$v_a\ll v_w$ particle population originating predominantly 
in the direction {\it normal} to the plane containing both 
${\bf v}_w$ and ${\bf B}$. 

Because of the relatively low particle velocities, their gravitational 
focusing is significant and can be approximated by the second 
term in the right hand side of equation (\ref{eq:F_ISS}). 
Combining equations (\ref{eq:flux_cold}), in which we neglect 
the term that is quadratic in $v/v_{w,\perp}$, (\ref{eq:F_ISS}), and 
(\ref{eq:dotQ}), we obtain the flux of particles of size
$r_g$ and velocity $v_a$ as
\ba
F^{\rm ISS}(v_a,&r_g,&r)=
\frac{3(4-\zeta)}{4\pi\rho r_{g,min}^4(r_0(v_a))}
\frac{dF_m(r_0(v_a),t)}{dV}
\nonumber \\
&\times&
\left[\frac{r_{g,min}(r_0(v_a))}{r_g}\right]^\zeta
\frac{GM_\odot}{r\Omega_L(r_g) v_{w,\perp}v_a},
\label{eq:FISS0}
\ea
where $r_0(v_a)$ is given by equation (\ref{eq:r_0_2}) and 
refers to the production site of particles in the Oort Cloud, 
while $r$ is the distance in the ISS at which 
they are detected.

In the limit $v_a\ll v_{w,\perp}$ equation (\ref{eq:r_0_2}) becomes
\ba
r_0(v_a)\approx \frac{R_L}{2}\left(\frac{v_a}{v_w}\right)^2,
\label{eq:low_v}
\ea
which is just the distance that a particle starting at rest and
moving at constant acceleration $F_{\rm E}/m_g$ needs to cover to reach 
velocity $v_a$. In the cold limit, the minimum velocity, 
$v_{min}$, that particles attain is set by $r_{in}$ via 
equations (\ref{eq:R_L}) and (\ref{eq:low_v}):
\ba
v_{min}&\approx& \left(\frac{3\sin\theta}{2\pi}
\frac{U v_w B r_0}{\rho c}\right)^{1/2}r_g^{-1}
\\
&\approx&
8~\mbox{km s}^{-1}\left(\frac{r_{in,3}U_1 B_5}{\rho_1 r_{g,1}^2}\frac{v_{w,\perp}}{15~{\rm km~s}^{-1}}\right)^{1/2}.
\label{eq:v_min}
\ea

The total flux of particles $F^{\rm ISS}(r_g,r)$ with velocities 
above $v_{min}$
can be found by integrating (\ref{eq:FISS0}) over velocities 
above $v_{min}$, which according to equation (\ref{eq:low_v})
is equivalent to integrating over $r_0$ starting at $r_{in}$,
and, effectively, to infinity since only a small fraction
of particles will be coming from $r$ significantly larger 
than $r_{in}$:
\ba
F^{\rm ISS}(r_g,r)&=&\frac{GM_\odot}{r\Omega_L v_{w,\perp}}
\int\limits_{r_{in}}^\infty \frac{3(4-\zeta)}
{8\pi\rho r_{g,min}^4(r_0)}
\nonumber \\
&\times&
\frac{dF_m(r_0,t)}{dV}
\left[\frac{r_{g,min}(r_0)}{r_g}\right]^\zeta
\frac{dr_0}{r_0}.
\label{eq:FISS}
\ea

We will now calculate the flux of unbound particles with
sizes between $r_{1}$ and $r_{2}$ originating
in the Oort Cloud, which requires additionally integrating equation 
(\ref{eq:FISS}) over $r_g$. For definiteness we take the power law 
indices of the Cloud mass distribution (equation \ref{eq:rho_0}) 
$\eta=3.5$, of the debris mass spectrum (equation \ref{eq:dotQ}) 
$\zeta=3.5$, and of the erosion time scaling (equation \ref{eq:ter_r}) 
$\kappa=4.5$ for this estimate. With these assumptions, equations 
(\ref{eq:Fm0}), (\ref{eq:d_cut}), and (\ref{eq:dotQ}) result in 
$\dot Q(r_0)\propto r_0^{-8.5}$ for $t\lesssim t_{er}(r_0)$.

Using equations (\ref{eq:R_L}), (\ref{eq:Fm0}), 
(\ref{eq:Fmev}) we find the following expression 
for the flux of particles with sizes $r_{1}<r_g<r_{2}$ 
at distance $r$ in the ISS:
\ba
F^{\rm ISS}(r_{1},r_{2}|r) = 
I_1\left(\frac{r_{2}}{r_{1}}\right)
I_2\left(\frac{t}{t_{er,in}}\right)
\nonumber \\
\times \frac{(4-\zeta)(\eta-3)}{8\pi}
\frac{c v_K^2(r)M_{\rm OC}}{t_{er,in}U B v_{w,\perp}}
\frac{r_1^{3-\zeta}}{r^{4-\zeta}_{g,min}(r_{in})r_{in}^3}
\label{eq:FISS_full}
\\
\approx 15~\mbox{m}^{-2}\mbox{yr}^{-1} 
I_1 I_2\frac{M_{\rm OC,10}}{r_{in,3}^{3.5}U_1^{1.5}B_5^{1.5}r_{1,1}^{0.5}}
\nonumber \\
\times
\left[\frac{v_K(r)}{30~\mbox{km s}^{-1}}\right]^2
\left(\frac{15~\mbox{km s}^{-1}}{v_{w,\perp}}\right)^{3/2}
\frac{4~\mbox{Gyr}}{t_{er,in}},
\label{eq:FISS_full_est}
\ea
where $v_K^2(r)=GM_\odot/r$, $r_{1,1}\equiv r_1/1\mu$m and we 
used $\eta=\zeta=3.5$, $\rho=1$ g cm$^{-3}$ in the numerical
 estimate. Factors $I_{1,2}$ are defined as
\ba
I_1(z) &\equiv & \int\limits_1^z x^{2-\zeta}dx=\frac{1-z^{3-\zeta}}{\zeta-3},
\label{eq:I_1}\\
I_2(z) &\equiv & \int\limits_1^\infty x^{\zeta-\eta-\kappa-5}
\left(1+zx^{-\kappa}\right)^{-2}dx.
\label{eq:I_2}
\ea
and are the only place where the details of the fragmentation 
cascade and Cloud structure (i.e. indices $\zeta$, $\eta$, 
and $\kappa$) enter our estimate.

For $\zeta = 3.5$, $r_{g,1} = 1 {\rm \mu m}$, and 
$r_{g,1} = 3 {\rm \mu m}$, as in our simulations, one 
finds $I_1 \approx 0.8$. Note that $I_2$ varies slowly as long as 
$t/t_{er,in}\lesssim 1$ but then goes down as 
$t/t_{er,in}$ becomes large. As a result, the largest value 
of $I_2$ is achieved when $t_{er,in}\gtrsim t$, but large 
$t_{er,in}$ reduces particle flux according to equation 
(\ref{eq:FISS_full_est}). Thus, the largest particle flux 
is reached for $t_{er,in}\sim t$, i.e. when the erosion timescale is
roughly equal to the Cloud age, and $I_2\sim 0.05-0.1$.

As equation (\ref{eq:I_2}) demonstrates, setting 
$t_{er,in}=4$ Gyr, roughly equal to the age of the Solar 
System, results in a flux of just several m$^{-2}$ yr$^{-1}$ 
$\mu$m-sized particles of cometary origin in the inner Solar 
System. These particles have velocities  
$\gtrsim v_{min}$ given by equation (\ref{eq:v_min}), and 
their arrival directions should be clustered in
the direction normal to the plane containing the ISM ${\bf B}$ 
field and the ISM wind velocity. Note that in deriving this 
estimate, we made a number of assumptions, 
which we discuss in \S \ref{sect:cav}.

\begin{figure}[htp]			
\begin{center}
\includegraphics[bb = 18 400 330 718,clip,width=\columnwidth]{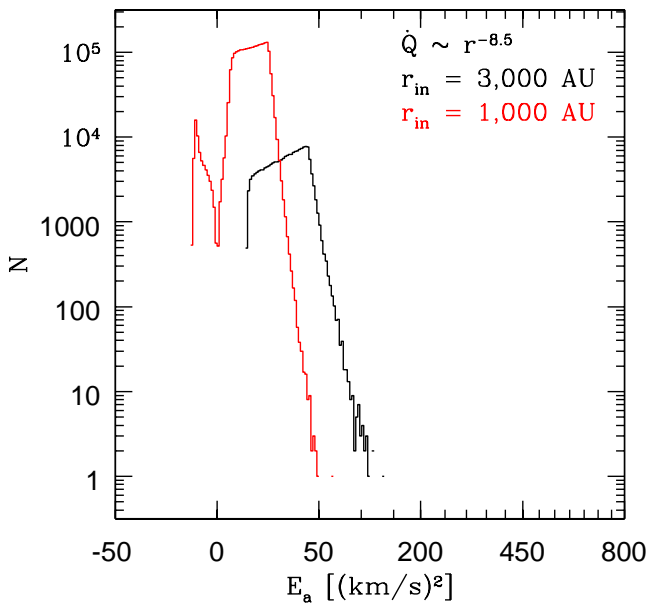}
\end{center}
\caption{Approach energy histograms for simulations of a plausible Oort Cloud structure with $\dot Q \propto r_0^{-8.5}$ and inner edges at 3,000 and 1,000 AU. Because of rapidly diminishing statistics, only particles from the innermost parts of the Cloud are ``observed'' in the inner Solar System, corresponding to lower approach specific energies (and velocities) than the other simulations. See the text for details.}
\label{E85}
\end{figure}

\begin{figure}[htp]			
\begin{center}
\subfigure{\includegraphics[bb = 14 50 828 527,clip,width=0.49\textwidth]{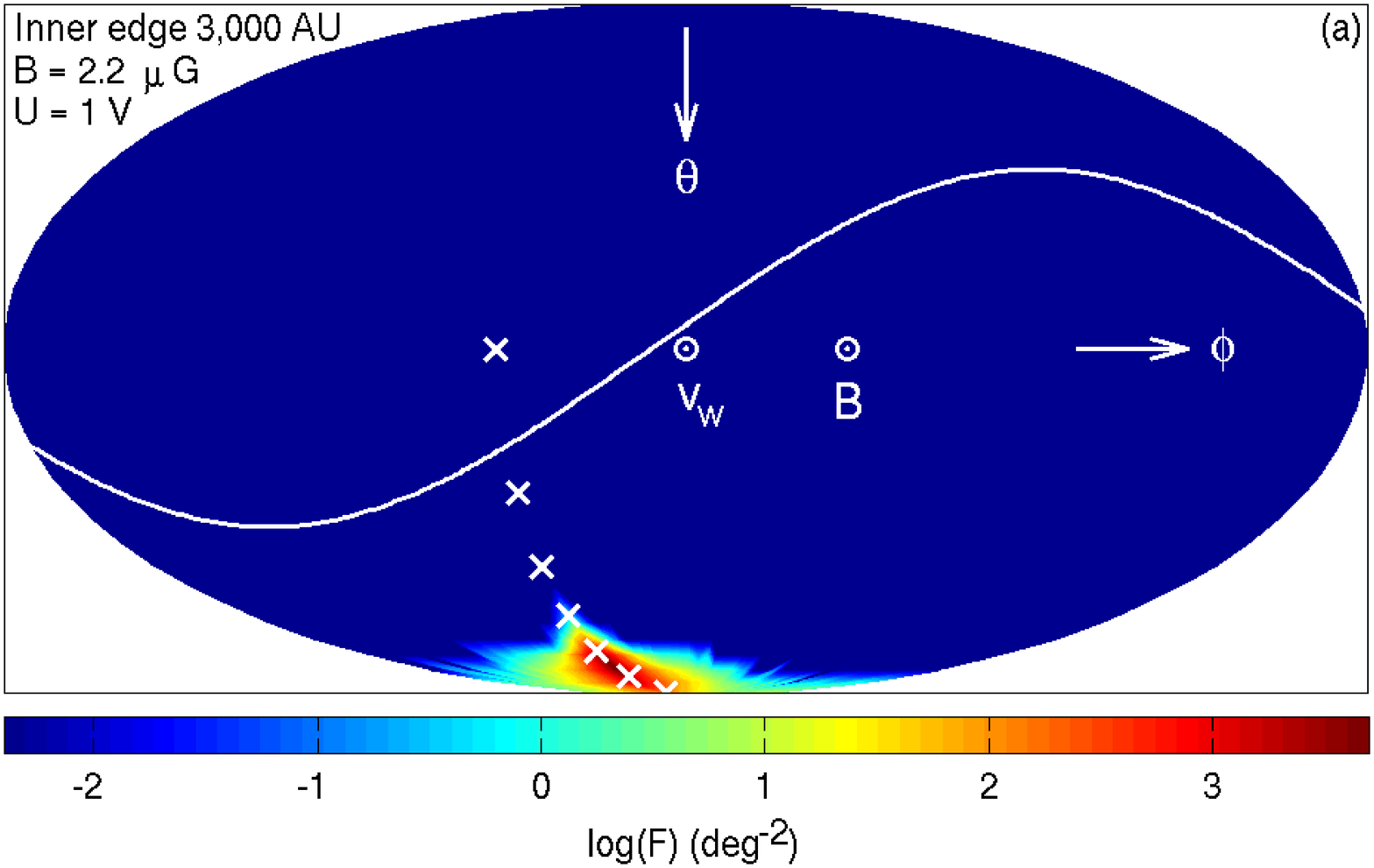}}
\subfigure{\includegraphics[bb = 14 14 828 527,clip,width=0.49\textwidth]{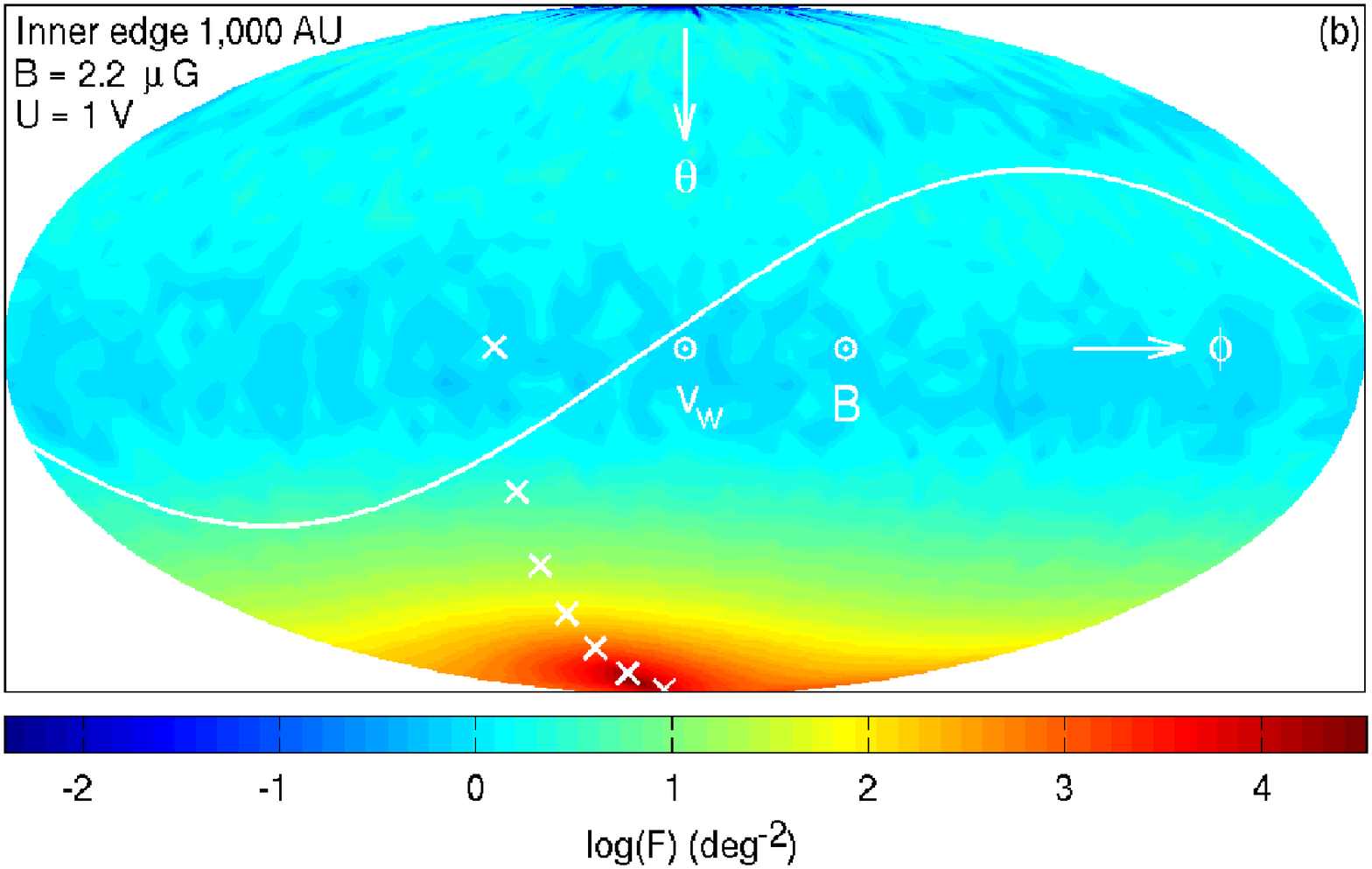}}
\end{center}
\caption{Maps of particle arrival directions for simulations of a plausible Oort Cloud structure with $\dot Q \propto r_0^{-8.5}$ and inner edges at 3,000 and 1,000 AU. Because of the steep production slope, only particles from the innermost parts of the Cloud are ``observed'' in the inner Solar System, arriving nearly perpendicular to both ${\bf v}_w$ and ${\bf B}$, i.e. at $\theta\approx 180^\circ$.}
\label{map85}
\end{figure}

\begin{figure}[htp]			
\begin{center}
\includegraphics[bb = 18 400 330 718,clip,width=\columnwidth]{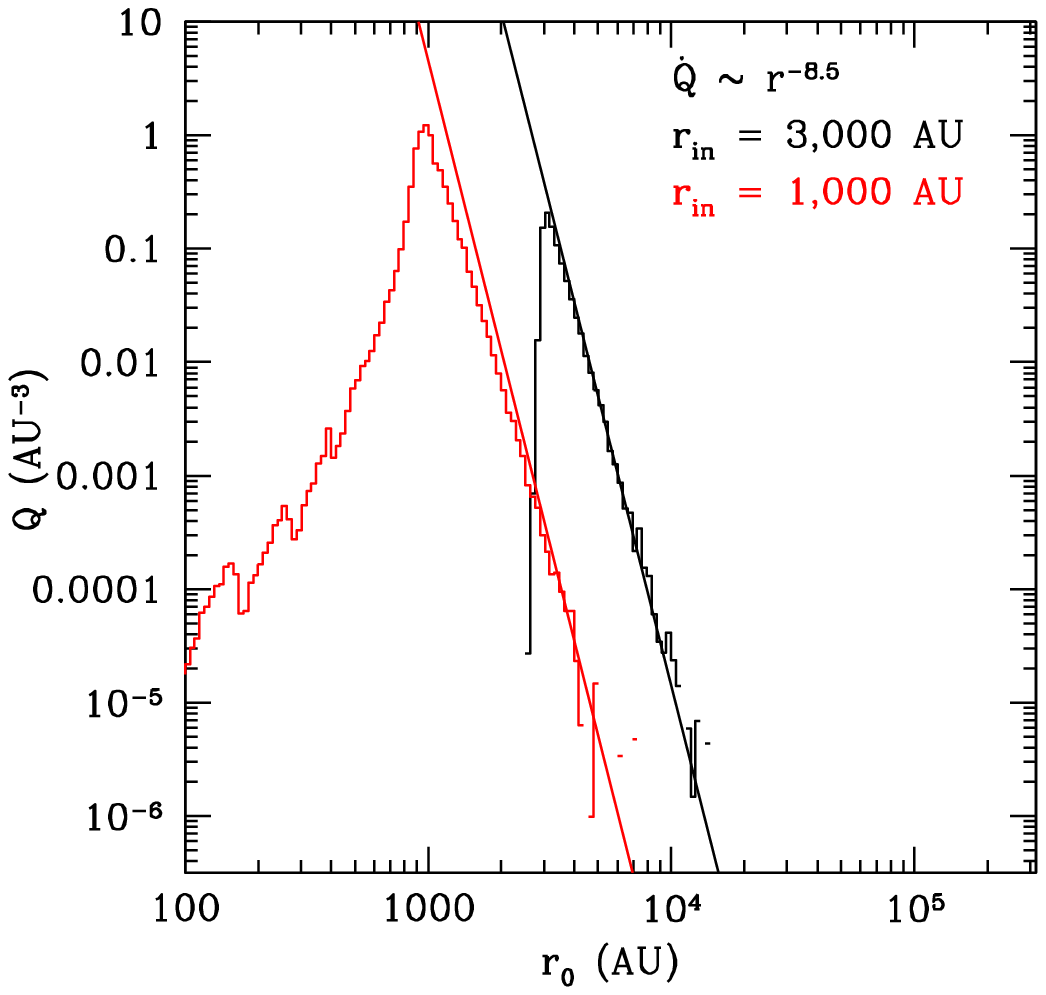}
\end{center}
\caption{Reconstructed $Q(r_0)$ (as a proxy for $\dot Q$) in simulations of a plausible Oort Cloud structure with $\dot Q \propto r^{-8.5}$, $\sigma = 0.7 v_K$, and inner edges at 3,000 and 1,000 AU. The slope of $\dot Q(r_0)$ is correctly recovered but because of its steepness there are not enough statistics to probe the structure of the distant parts of the Cloud.}
\label{r85}
\end{figure}


\subsection{Simulations of the expected Oort Cloud parameters}
\label{finalsims}

We now explore observable signatures of dust production in a particular Cloud model motivated by the results of this section. We assume $\dot Q(r_0)\propto r_0^{\zeta-(4+\eta+\kappa)}\propto r_0^{-8.5}$, which results from $\eta=\zeta=3.5$, $\kappa=4.5$. All other parameters are kept as in \S \ref{numerical}. 

Figure \ref{E85} shows that such a steep $\dot Q(r_0)$ profile results in particle velocity distribution in the ISS that is very heavily skewed toward low approach velocities. In the $r_{in}=3,000$ AU case, the peak of the distribution is at 10 km s$^{-1}$, and by $v_a=15$ km s$^{-1}$, the number of particles drops by about four orders of magnitude. About $8\times 10^{-6}$ of the total number of particles pass between 3 and 10 AU. This fraction increases to $2.4\times 10^{-4}$ for $r_{in}=10^3$ AU, and $\approx 2.7\%$ of these grains end up being bound to the Sun. The flux ratio for the two values of $r_{in}$ is about 30 and is somewhat less than $3^{3.5}\approx 47$ predicted by equation (\ref{eq:FISS_full_est}) --- an effect that we saw in $\dot Q(r_0)\propto r_0^{-4}$ case as well, see \S \ref{vel_dist}.

Maps of the particle arrival directions presented in Figure \ref{map85} show that essentially all unbound grains arrive from $\theta=180^\circ$. In the $r_{in}=10^3$ AU case, there is a roughly uniform background of bound particles, but the concentration of unbound particles far from both the ecliptic and ${\bf v}_w$ direction is still clearly seen.

Figure \ref{r85} shows that with a steep particle production profile, reconstructing the radial structure of the Cloud far from $r_{in}$ is not realistic. Only the innermost regions close to the inner edge can be probed, but over there, both the slope of $\dot Q(r_0)$ and the location of $r_{in}$ are recovered quite reliably by our reconstruction procedure.


\section{Discussion}
\label{sect:disc}


Our results show that cometary dust particles produced in the OSS
can be used as probes of the Oort Cloud's properties. 
Current or near-future 
technology should allow their detection, provided that the 
Cloud's properties are compatible with those used in evaluating 
particle flux in equation (\ref{eq:FISS_full_est}), i.e. a relatively 
close-in inner edge of the Cloud $r_{in}\sim 10^3$ AU, Cloud mass 
$M_{\rm OC}\sim 10M_\oplus$, and an erosion time at the inner edge 
$t_{er}(r_{in})$ comparable to the Solar System age of $4.5$ Gyr.
In this case, one expects a space-borne experiment to detect a flux of 
${\rm \mu}$m-sized cometary particles of several m$^{-2}$ yr$^{-1}$.

Dust detectors onboard {\it Galileo}, {\it Ulysses}, etc. have
already been able to measure fluxes of ${\rm \mu m}$-sized grains moving 
at $20-30$ km s$^{-1}$ at the level of tens of m$^{-2}$ yr$^{-1}$, see 
Table \ref{dust}. Thus, we expect the next generation of dust 
detectors \citep{DUNE,Grun12} to be able to measure the predicted cometary 
particle fluxes. Observationally, these particles can hardly 
be confused with interplanetary or interstellar dust grains  
as the latter should be concentrated toward the ecliptic and 
the ISM wind direction, correspondingly. At the same time, the 
cometary dust particles, even the relatively slow ones with 
$v_a\lesssim 10$ km s$^{-1}$, have a very distinct clustering of 
approach directions, orthogonal to the plane containing ${\bf v}_w$ 
and ${\bf B}$, see \S \ref{vel_dir}.

Detection of the cometary grain population in the ISS would strongly
imply that the inner edge of the Oort Cloud lies at a relatively small 
separation, around $10^3$ AU. This is not inconceivable given the 
existence of OSS objects such as Sedna \citep{Brown}, the orbit 
of which, given its large perihelion (76.4 AU) and relatively small 
semi-major axis (519 AU), can hardly be explained in the framework 
of the standard picture of scattered Kuiper Belt object origins.  
If Sedna is in fact a member of the inner extension of the Oort Cloud,
then $r_{in}\sim 10^3$ AU should be possible. 

Larger values of $r_{in}$ can make cometary grains produced in the 
Oort Cloud itself undetectable, given the strong dependence of the
particle flux on $r_{in}$, see equation (\ref{eq:FISS_full}). Indeed, 
moving $r_{in}$ to 3,000 AU would result in the reduction of $F^{\rm ISS}$
by a factor of 80, likely rendering cometary dust unobservable. 

Our flux estimate (\ref{eq:FISS_full}) also strongly relies on a very 
uncertain erosion time of the Cloud. The best chances of seeing 
cometary dust are for $t_{er}$ similar to the Solar System age. 
Shorter $t_{er}$ means that the current Cloud mass and dust production 
in it are reduced by collisional grinding compared with the original 
$M_{\rm OC}$. Longer $t_{er}$ lowers the mass flux in the collisional 
cascade, see equation (\ref{eq:FISS_full}). Thus, a certain degree of 
fine-tuning of Cloud properties is needed for $F^{\rm ISS}$ to be measurable. 
On the other hand, our estimates of $t_{er}$ in Appendix 
\ref{sect:erosion} suggest that $t_{er}$ of order several Gyr at $r_{in} \sim 10^3$ AU is 
plausible if comets are internally weak bodies, as they likely are.

As we showed in \S \ref{sect:part_flux} and \S \ref{Qdot}, particle 
velocity distribution contains important information about both the 
Cloud characteristics and the ISM wind properties. The maximum 
velocity of the cometary grains is a direct measure of the wind 
velocity projected on the ISM magnetic field direction, $v_{w,\perp}$. 
It is important to note that cometary particle trajectories probe 
the {\it global} wind structure on large, $10^4$ AU ($\sim R_L$) 
scales, as the effect of small-scale deviations from the uniform flow 
permeated by a homogeneous $B$-field on the particle motion likely 
averages out. The particle arrival direction peak gives the orientation 
of ${\bf B}$ in the sky plane, perpendicular to ${\bf v}_w$. 
At the same time, the existing methods of inferring 
$v_w$, $B$ and $\theta_{wB}$ (see references in Table \ref{ISMmodels}) probe only the {\it local} wind 
structure, since they rely on information gathered on heliospheric 
scales of order $10^2$ AU. Thus, cometary grains can give us access 
to the global view of the ISM in the Solar vicinity, and it should 
not be surprising if the inferred value of $v_{w,\perp}$ is 
different from the locally measured one. 
 
The slope of the grain velocity distribution carries information about 
the distribution of the dust production rate $\dot Q$ in the OSS 
(\S \ref{Qdot}). The latter, in turn, can be related to the most important 
underlying characteristics of the Cloud--the spatial distribution 
of the cometary density, the total mass of the Cloud, the location of its 
inner edge, and its erosion timescale, see equation (\ref{eq:dotQ1}). 
The latter characteristic is itself a function of the others and of 
the physical properties of comets (their internal strength).

Needless to say, extracting all this information from the behavior
of the particle velocity distribution requires a very well sampled $F(v_a)$,
which will likely be very difficult, even with the next generation of 
dust detectors. As Figure \ref{E85} clearly shows, probing the high-$v_a$
tail of $F(v_a)$, which provides information on $v_{w,\perp}$,  
for realistic (i.e. steeply declining with $r_0$) behaviors of 
$\dot Q(r_0)$ would require detectors with very large collecting 
areas, given the low expected fluxes. Even probing the low-$v_a$ 
slope of $F(v_a)$ to infer $\dot Q(r_0)$ in the inner 
Cloud would require at least 
several $10^2-10^3$ detected particles, see Figures \ref{E85} \&
\ref{r85}. If the overall grain flux $F^{\rm ISS}$ turns out to be 
too low for the reasons described previously, this task will be infeasible.

An added complication is that the interpretation of possible 
measurements of $F_{\rm ISS}$ and velocity distribution suffers from a 
number of degeneracies, as equations (\ref{eq:Qr}) and 
(\ref{eq:FISS_full}) demonstrate. For example, the derivation of 
$\dot Q(r_0)$ via equation (\ref{eq:Qr}) depends on the assumed 
$B$-field strength, since both $\Omega_L$ and $R_L$ in equation 
(\ref{eq:r_0_2}) depend on $B$. Overestimating $B$ by a factor of $\lambda$
would result in underestimating all the linear scales, including 
$r_{in}$, by a factor of $\lambda$, and overestimating $\dot Q$
by the same factor. The overall dust production rate in the Cloud
$\sim \dot Q(r_{in})r_{in}^3$ would be underestimated by $\lambda^2$.
On the other hand, this particular degeneracy might be broken if  
one can accurately sample $F(v_a)$ for particles of different 
sizes, as $\Omega_L$ and $R_L$ are functions 
of $r_g$, while the spatial distribution of $\dot Q$ is 
independent of particle size.


\subsection{Comparison with other detection methods}
\label{sect:compare}

A collisional cascade in the Oort Cloud can manifest itself not only via dust particles passing through the ISS, but also small dust with sizes down to $10{\rm \mu}$m can reveal itself via (1) back-scattering of Solar light, (2) thermal radiation, and (3) gamma-ray emission produced in cosmic ray collisions with sub-meter-sized objects.  

We have verified that our Cloud model motivated by the calculation of its collisional evolution and explored in \S \ref{subsect:coll} is not a strong contributor to the optical and near-IR sky background. For a Cloud with mass $10M_\oplus$ and $r_{in}=10^3$ AU, the scattered light intensity in the ISS is about $10^{-11}$ ergs s$^{-1}$ cm$^{-2}$ sr$^{-1}$ $\AA^{-1}$ at $\lambda\sim 1\mu$m, even for a particle albedo of unity. This is about four orders of magnitude less than that produced by zodiacal dust, which is about $7\times 10^{-8}$ ergs s$^{-1}$ cm$^{-2}$ sr$^{-1}$ $\AA^{-1}$ at $800$ nm, see Bernstein et al (2002). The scattered light intensity varies roughly as $r_{in}^{-4.7}$ and becomes comparable to the zodiacal light only for $r_{in}\lesssim 10^2$ AU.

Babich \etal (2007) explored ways of constraining the Oort Cloud's properties using thermal emission from its small constituent particles. They have shown that a $10M_\oplus$ Cloud with $r_{in}=10^3$ AU is below the threshold of current CMB experiments meaning that our favored Cloud model is undetectable via its own thermal radiation. 

Moskalenko \& Porter (2009) investigated cosmic ray induced $\gamma$-ray production in the Oort Cloud. They showed that the Cloud becomes a noticeable contributor to the $\gamma$-ray background only if its mass in objects with $r_g\lesssim 1$ m (most efficient at producing $\gamma$-rays) is $\gtrsim 50M_\oplus$ for $r_{in}=10^3$ AU (this number varies as $r_{in}^2$). The mass in objects of all sizes up to $d_m\sim 1$ km would be $\sim 10$ times higher. Our typical model with $M_{\rm OC}=10M_\oplus$ is far below this limit. 

We can thus conclude that other ways of directly probing the Cloud properties explored in the literature cannot reveal its presence if $M_{\rm OC}\lesssim 10M_\oplus$ and $r_{in}\gtrsim 10^3$ AU. Direct detection of cometary dust particles originating in the OSS may be our only hope for that.


\subsection{Contribution to zodiacal dust}
\label{sect:zodi}

According to Figure \ref{Esigma}, some particles produced in collisions in the Oort Cloud have negative specific energy when passing through the ISS, i.e. they are bound to the Sun. Inside the heliosphere, their orbits may be additionally affected by the electromagnetic effects of the Solar wind. These particles would be contributing to the zodiacal dust population in the ISS. Thus, cometary collisions in the OSS might provide a previously unrecognized source of the isotropic component of zodiacal dust. 

In this work, we do not pursue a detailed assessment of the zodiacal dust production rate via this channel but merely mention some of its characteristics. First, bound dust particles are produced with non-zero efficiency only when the inner edge of the Cloud lies close to the Sun, see Figures \ref{E4}a \& \ref{Esigma}. As discussed in \S \ref{vel_dist}, for $r_{in}=10^3$ AU and $\sigma_v=0.7v_K$, bound particles passing between 3-10 AU represent $\sim 10^{-6}$ of the total number of grains created in the Oort Cloud (about 1.7\% of all particles passing through this annulus), while already for $r_{in}=3,000$ AU, essentially all such grains are unbound. 

Second, Figure \ref{Esigma} shows that the number of bound particles is a strong function of their velocity dispersion at birth: for $\sigma_v=0.7v_K$, the flux of bound particles is $\sim 15$ times higher than for $\sigma_v=0.3v_K$. Thus, properly accounting for the initial grain velocities is important for understanding the production of bound particles. 

Even though bound particles represent only a small fraction of the dust generated in the Oort Cloud, the Cloud mass may be sufficiently high for these grains to be a non-negligible contributor to the zodiacal dust population. We leave a detailed investigation of this issue to future work. 

\begin{figure}[htp]			
\begin{center}
\subfigure{\includegraphics[bb = 14 50 828 527,clip,width=0.49\textwidth]{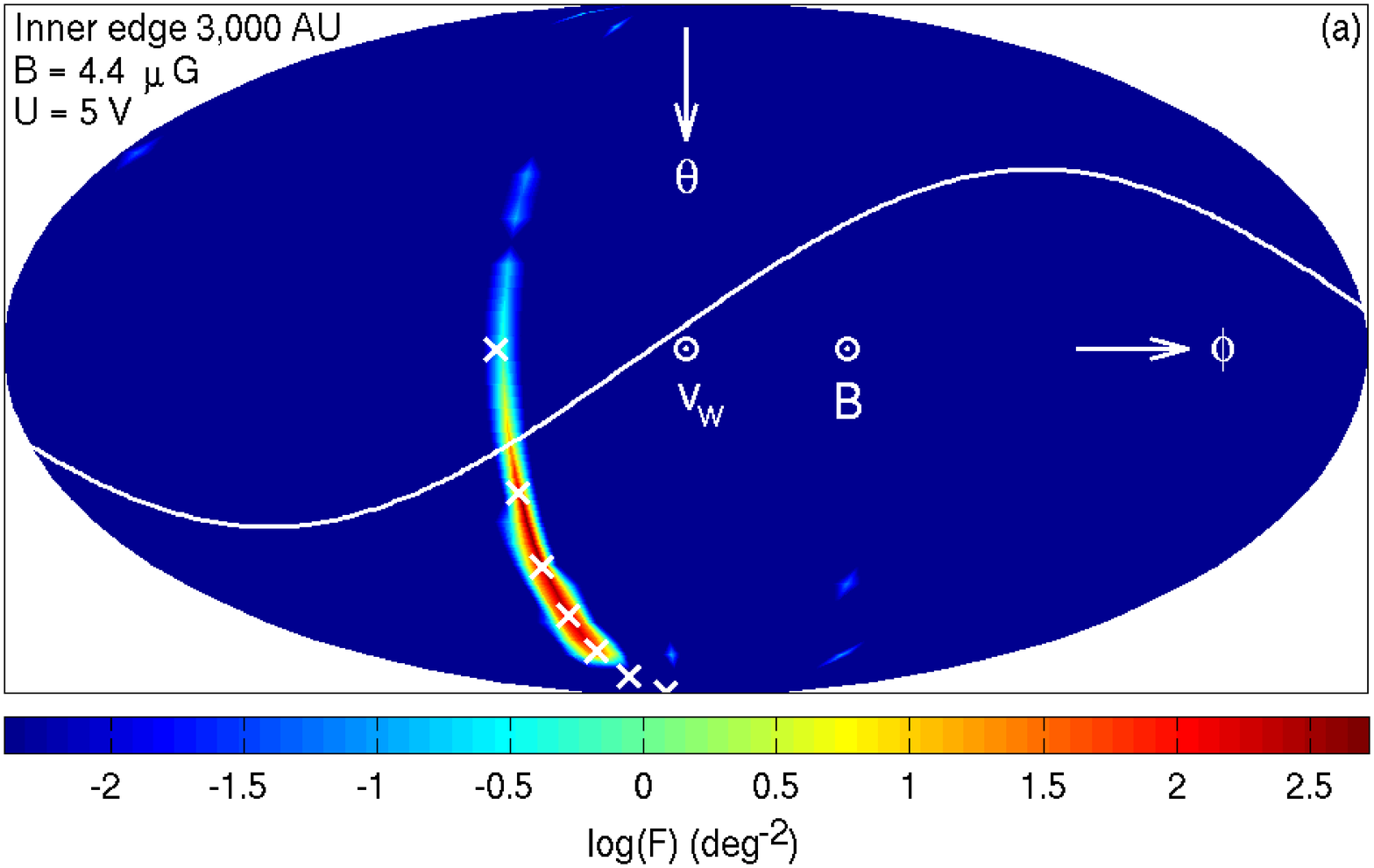}}
\subfigure{\includegraphics[bb = 14 14 828 527,clip,width=0.49\textwidth]{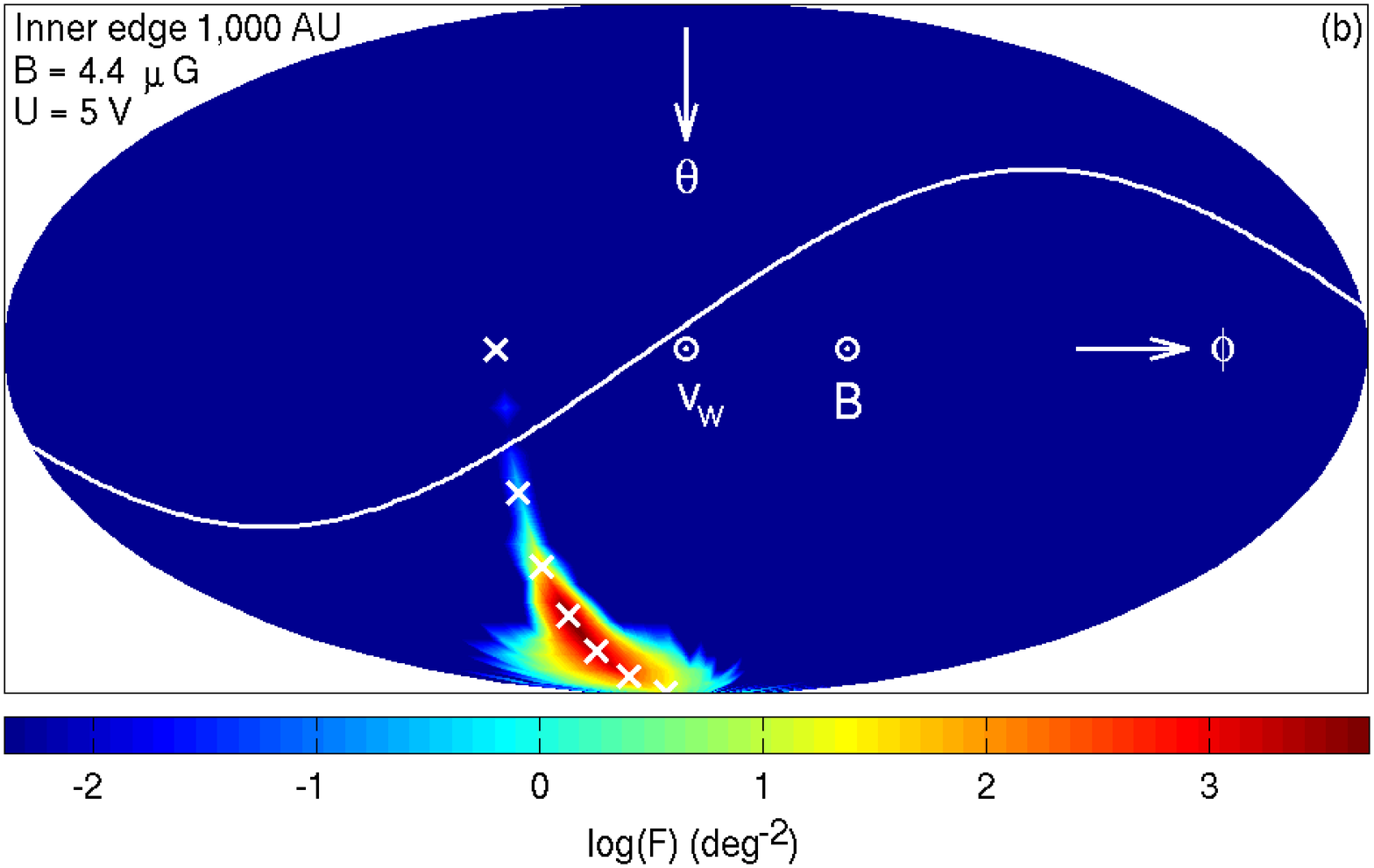}}
\end{center}
\caption{Maps of particle arrival directions for simulations of a plausible Oort Cloud structure with $\dot Q \sim r^{-8.5}$ and a larger electric potential (5 V) and magnetic field (4.4 ${\rm \mu}$G). The larger electromagnetic forces result in particle arrival directions shifting closer to the wind direction.}
\label{map5V85}
\end{figure}

\begin{figure}[htp]			
\begin{center}
\includegraphics[bb = 18 144 330 450,clip,width=\columnwidth]{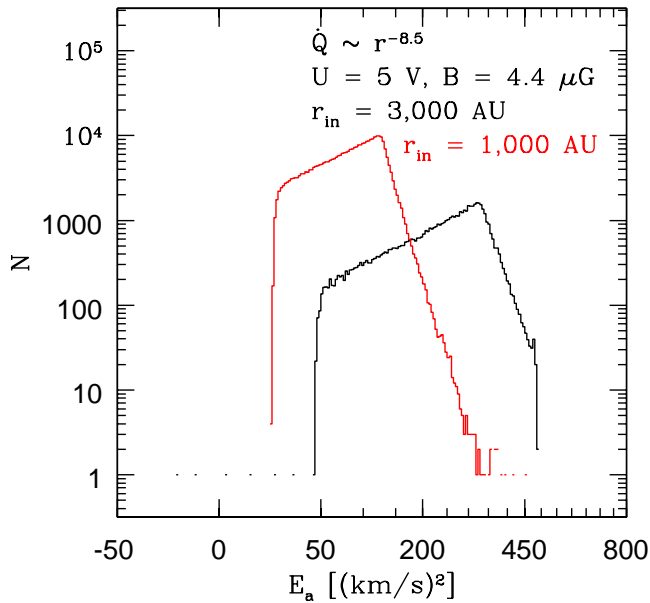}
\end{center}
\caption{Approach energy histograms for simulations of a plausible Oort Cloud structure with $\dot Q \sim r^{-8.5}$ and a larger electric potential (5 V) and magnetic field (4.4 ${\rm \mu}$G). This setup results in higher approach velocities (and energies) consistent with the value of $v_w$.}
\label{E5V85}
\end{figure}


\subsection{Origin of big ISM particles}
\label{sect:appl2}

\citet{Frisch} and \citet{Landgraf} noted the existence of a 
significant population of big, ${\rm \mu m}$-sized grains (with a 
flux of tens of m$^{-2}$ yr$^{-1}$, see Table \ref{dust}), arriving 
in the general direction of the ISM wind flow with speeds comparable 
to $v_w$. The natural interpretation of such grains as being  
a part of the local ISM is very problematic, as shown by \citet{Draineconundrum}. 

\citet{Frisch} looked into an alternative 
origin of this population, namely cometary dust produced in the 
Oort Cloud. By placing the dust source as far as $5\times 10^4$ AU
away from the Sun and assuming all big particles to have a cometary 
origin, they inferred a very short erosion timescale of the Cloud, less than 
$10^5$ yr, which is unrealistic. For this reason \citet{Frisch} 
dismissed a cometary origin for the ${\rm \mu m}$-sized interstellar dust 
particles.

Our results allow us to reassess this problem. Explaining the 
flux of ${\rm \mu m}$-sized particles by cometary grains may in fact be 
marginally possible, contrary to the \citet{Frisch} estimates. 
The main difference is that most of the Cloud mass is likely 
concentrated much closer to the Sun than the $5\times 10^4$ AU 
adopted in that study. As our equation (\ref{eq:FISS_full_est}) 
shows, this tremendously boosts the particle flux in the ISS 
for the erosion timescale of order the Solar System age. Our
more careful calculation of the particle flux in the ISS 
presented in \S \ref{sect:flux_ISS} still shows that $F^{\rm ISS}$ 
falls short of explaining the flux of big interstellar particles 
by about an order of magnitude, but the discrepancy is not at the
level of $10^5$ as in \citet{Frisch}.

By exploring the particle {\it delivery} issue, not addressed in 
\citet{Frisch} (they essentially assume that cometary dust 
couples effectively to the ISM wind), we also show that particles 
get accelerated to speeds of order $v_w$ (and even higher) and 
are coupled to the ISM flow on scales of order $R_L\sim 10^4$ AU.
However, another problem emerges: we generically find that the 
majority of the cometary particles arrive to the ISS in 
the direction {\it orthogonal} to
the ISM wind direction, see \S \ref{vel_dir}. This is very hard 
to reconcile with observations. 

We have attempted to find the conditions under which cometary 
particles would arrive to the ISS parallel to ${\bf v}_w$. This
can be accomplished by increasing the curvature of the particle 
trajectory in the Oort Cloud, e.g. due to a stronger $B$-field and/or
higher grain potential $U$. In Figure \ref{map5V85}, we show the distribution 
of particle arrival directions obtained for $U=5$ V and $B=4.4$ ${\rm \mu}$G
(so that $UB$ is 10 times higher than our normal value) and 
$r_{in}=3\times 10^3$ AU. One can see that the locus of arrival 
directions gets displaced toward that of the interstellar wind. 
However, a brief look at equation (\ref{eq:FISS_full_est}) 
demonstrates that for this choice of parameters, the particle 
flux is greatly reduced ($10^{-2} - 10^{-3}$ m$^{-2}$ yr$^{-1}$) 
and is much less than the value for big ISM dust grains. 

Similarly, Figure \ref{E5V85} shows the velocity histograms for these simulations with stronger electric and magnetic fields. Since particles are accelerated to high speeds by electromagnetic forces much more rapidly in these cases, the overall distribution is one of higher approach energies. In the $r_{in} = 3,000$ AU case, observable fluxes of particles occur up to the velocity limit of $2v_{w,\perp}$ but are generally quite low.

Thus, we find that simultaneously reconciling the observed flux 
and direction of fast ${\rm \mu m}$-sized grains with their possible 
cometary origin is very difficult. For that reason we also 
find it {\it unlikely} that these big grains can be produced in 
cometary collisions in the Oort Cloud.

On the other hand, we show that most of the $(1-10)\mu$m particles produced in the Oort Cloud are swept out of the Solar System by the ISM wind. Depending on the lifetime of large, icy grains in the ISM and the production rate of such grains around other stars, it may be possible for a global population of large grains to form in the ISM from extrasolar cometary material. These interstellar grains would ultimately be coupled to the ISM gas, making them a candidate for the observed population of large ISM particles. We leave the investigation of this potential source of interstellar grains to future work.


\subsection{Possible extensions and limitations of our calculation}
\label{sect:cav}

Our analysis of cometary grain dynamics and their collisional 
origin has assumed a number of simplifications that we list below 
and that can be easily relaxed in the future.

Our calculation of the particle trajectories in both \S 
\ref{sect:small_dyn} and \S \ref{numerical} assumed that the ISM 
properties are uniform on scales of $\sim 10^4-10^5$ AU. This 
may not be the case if the local ISM is turbulent on these scales. This 
can easily be accounted for by introducing a random component 
of ${\bf B}$ and ${\bf v}_w$ in our numerical calculations. 
However, we expect these effects to not be very important 
because of averaging on larger scales.

The grain potential $U$ should increase toward the Sun. We have also 
neglected the influence of the Solar wind on the grain motion 
inside the heliosphere. However, (1) this is going to be important 
only on scales $\sim 10^2$ AU $\ll R_L$, and (2) the Solar wind 
effect can be corrected for when the satellite measurements 
are made \citep{Landgraf1}. Given the uncertainties of the model, 
we do not feel that inclusion of these effects is justified. 

Our flux calculation in \S \ref{sect:dust_prod} is very 
dependent on input parameters, as equation (\ref{eq:FISS_full}) 
demonstrates. The biggest uncertainty, qualitatively affecting 
the results of this study, is the location of the inner edge of
the Cloud $r_{in}$. Related to that is our assumption of the 
spherically-symmetric Oort Cloud: galactic tides are 
likely to render its structure triaxial, and models suggest that 
it made be flattened in the inner regions \citep{Dones}. Both of 
these effects could influence the dominant particle arrival 
direction and could introduce systematic errors into our 
ISS particle flux calculations.

The erosion timescale of the Cloud at its inner edge is another poorly 
constrained parameter to which the particle flux calculation is
very sensitive. Its calculation involves not only knowing 
the local density of comets, but also understanding their internal 
properties, such as the collisional strength, see Appendix 
\ref{sect:erosion}. For this reason the value of $t_{er}$ is 
at present highly uncertain (by $\sim 2$ orders of magnitude). 

Finally, the Cloud mass $M_{\rm OC}$ has been only weakly 
constrained by observations \citep{Dones} and is also a free parameter 
in our flux estimates, see \S \ref{sect:dust_prod}.


\section{Summary}
\label{sect:concl}

We have investigated the possibility of probing the properties of the
Oort Cloud via the direct detection of dust grains produced in collisions
of comets in the OSS. We explored both the production of cometary dust
in the Cloud and its delivery to the ISS over the $\sim 10^4$ AU scales
under the influence of the electromagnetic forces produced by the magnetized 
interstellar wind. 

We present an analytic model of grain dynamics, which 
allows us to relate the details of dust production inside the Oort Cloud 
to measurable dust observables such as the velocity distribution 
of dust grains. We demonstrate that the latter can be used to reconstruct
the spatial distribution of the dust production profile in the OSS and to
infer important information about the ISM wind properties on large scales,
$\sim 10^4$ AU, given good enough statistics. We show that cometary grains
arrive to the ISS predominantly in the direction orthogonal to
both the ISM wind speed ${\bf v}_w$ and the ISM ${\bf B}$-field direction.
The majority of these particles should have approach speeds of $\lesssim 10$ 
km s$^{-1}$. These kinematic properties distinguish cometary grains from both 
interplanetary and interstellar dust grains. These predictions were 
successfully verified with direct numerical simulations of dust grain 
trajectories under the effect of the ISM wind.

We have also developed a simple model for the collisional evolution of
the Oort Cloud and used it to compute the flux of cometary grains in
the inner Solar System. We predict a flux of several m$^{-2}$ yr$^{-1}$ for 
$\sim (1-3) {\rm \mu}$m particles for a Cloud mass of $10$ M$_\oplus$, inner edge 
$r_{in} = 10^3$ AU, and the erosion time at the inner edge 
$t_{er}(r_{in})\sim 5$ Gyr. We view the latter two parameters as the most 
uncertain inputs of our model. A significantly different value of 
$t_{er}(r_{in})$ or larger value of $r_{in}$ would likely make the cometary 
dust population undetectable. 

We also explore the possibility of explaining the recently detected
population of big, ${\rm \mu m}$-sized grains entering the ISS with the ISM flow
via their cometary origin. By combining the kinematic 
and flux constraints for the cometary grains, we find this explanation to be 
highly unlikely.

Based on our results, we claim that a possible detection of cometary 
dust grains entering the ISS should provide us with an independent 
probe of the properties of both the Oort Cloud and the properties of 
the local ISM, and we encourage efforts to build next-generation dust 
detectors for this purpose. 


\acknowledgements

We are indebted to Bruce Draine for instigating our interest
in this problem, continuous encouragement, numerous discussions, 
and advice. 


\appendix

\section{Description and Tests of Simulations}
\label{description}

\begin{figure*}[htp]			
\begin{center}
\subfigure{\includegraphics[bb = 14 50 828 527,clip,width=0.49\textwidth]{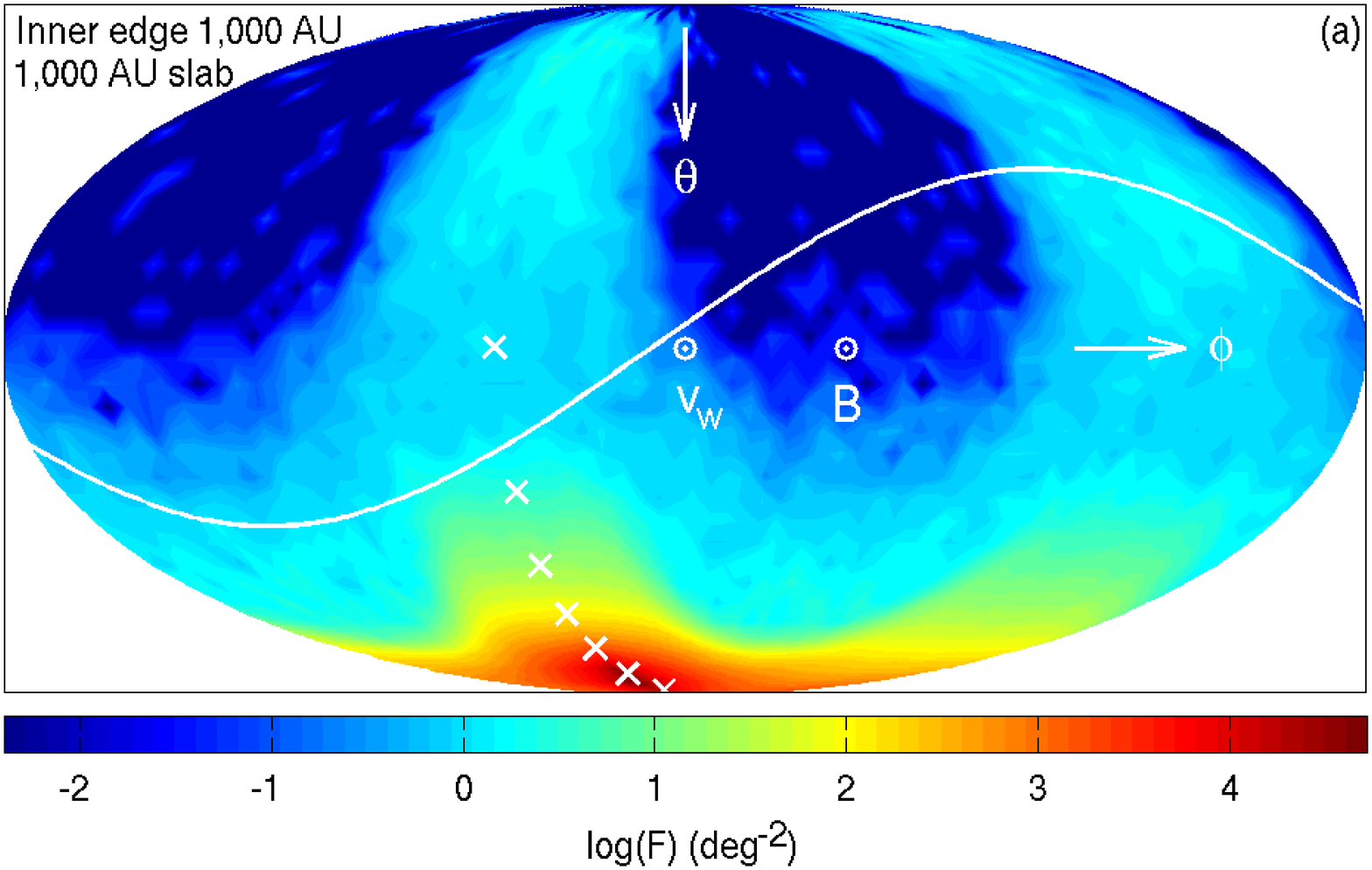}}
\subfigure{\includegraphics[bb = 14 50 828 527,clip,width=0.49\textwidth]{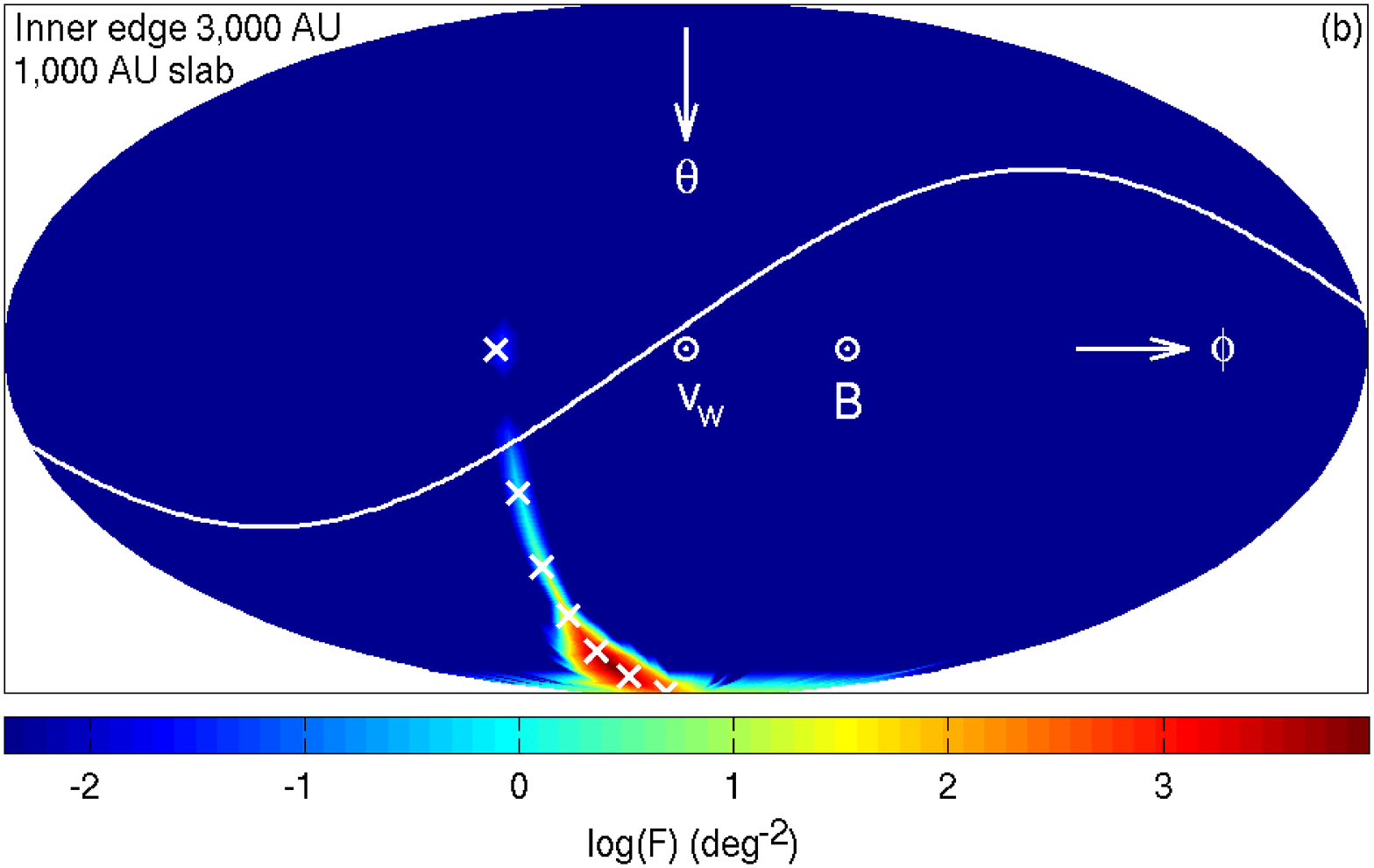}}
\subfigure{\includegraphics[bb = 14 50 828 527,clip,width=0.49\textwidth]{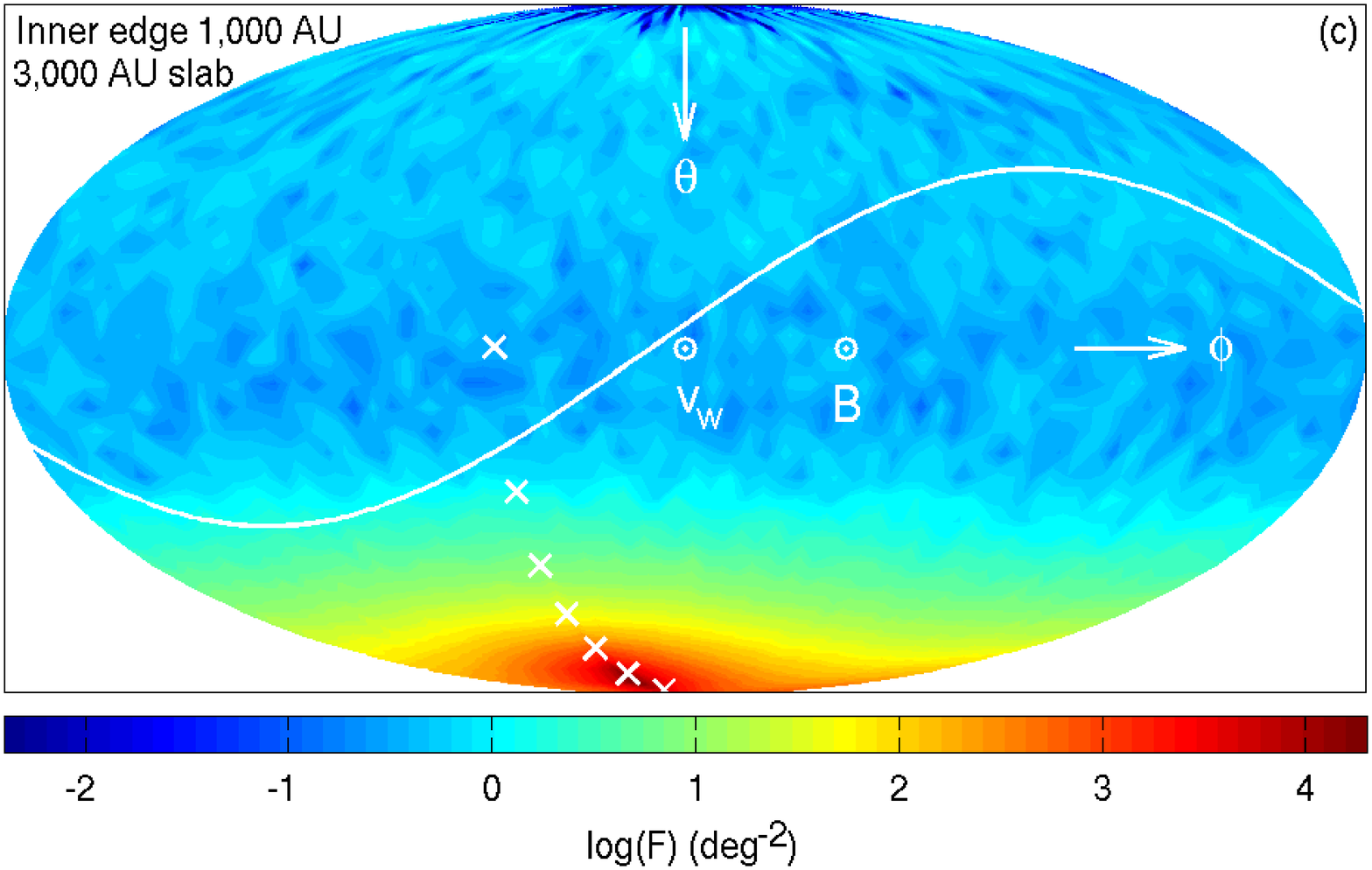}}
\subfigure{\includegraphics[bb = 14 50 828 527,clip,width=0.49\textwidth]{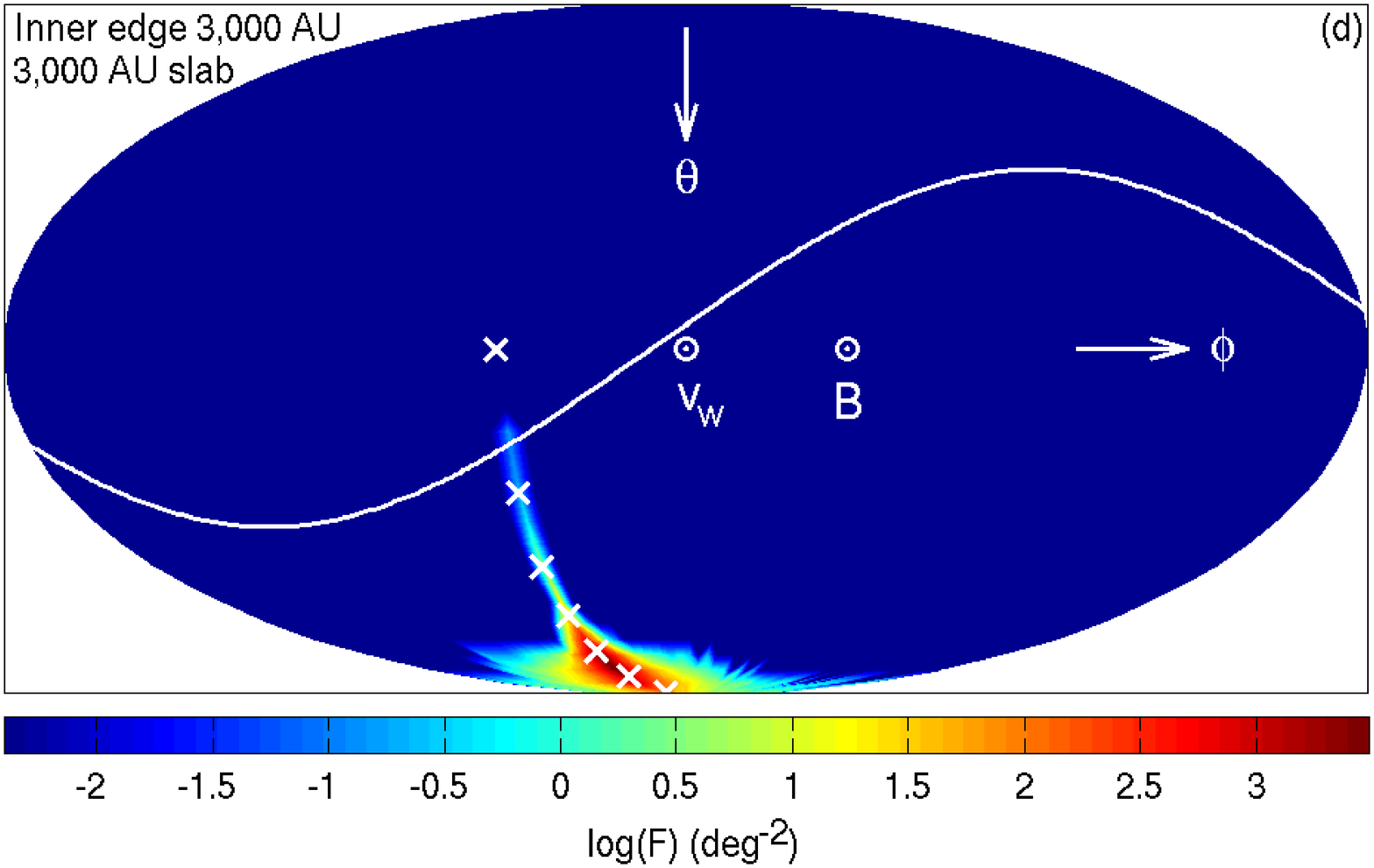}}
\subfigure{\includegraphics[bb = 14 14 828 527,clip,width=0.49\textwidth]{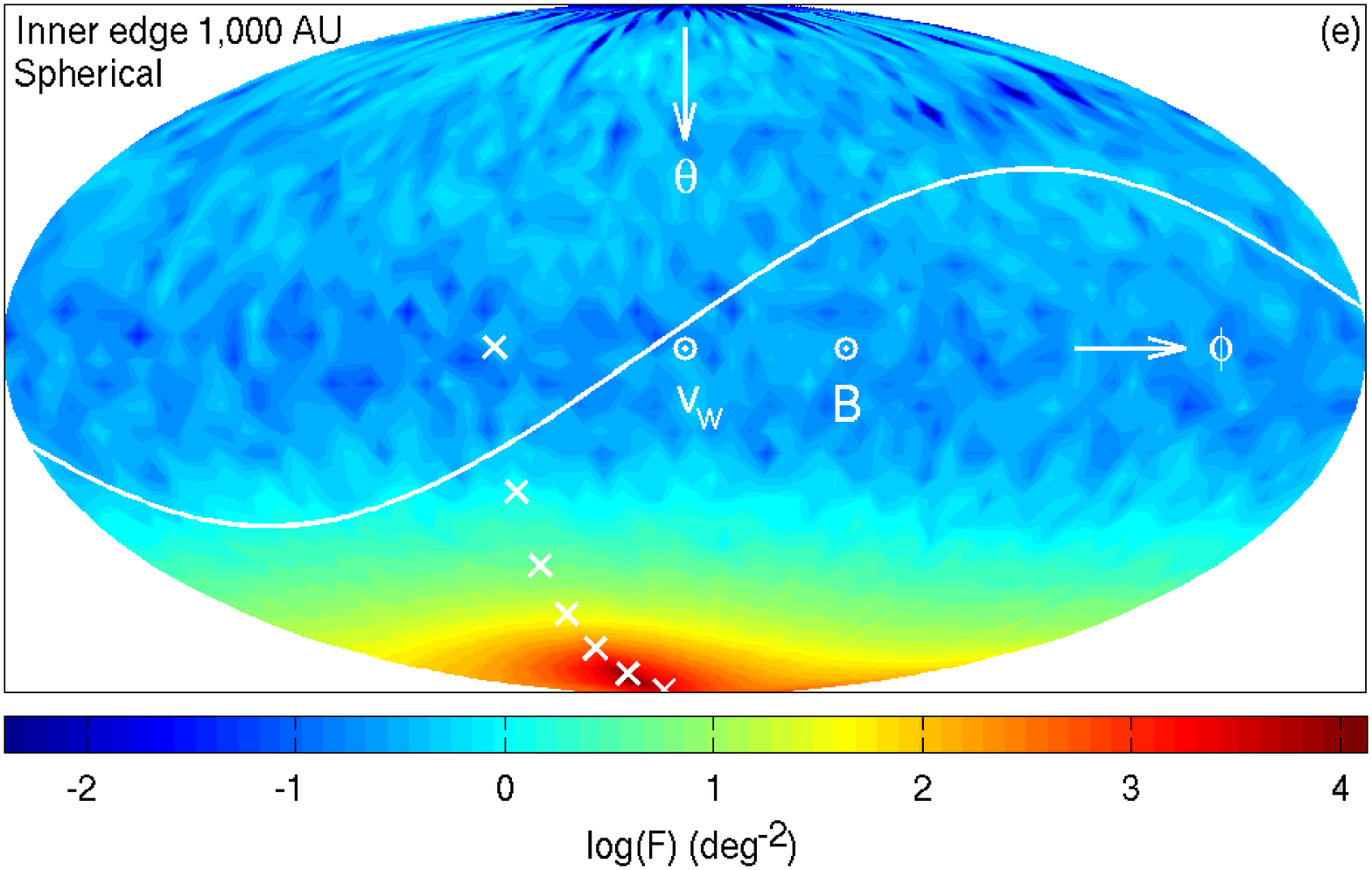}}
\subfigure{\includegraphics[bb = 14 14 828 527,clip,width=0.49\textwidth]{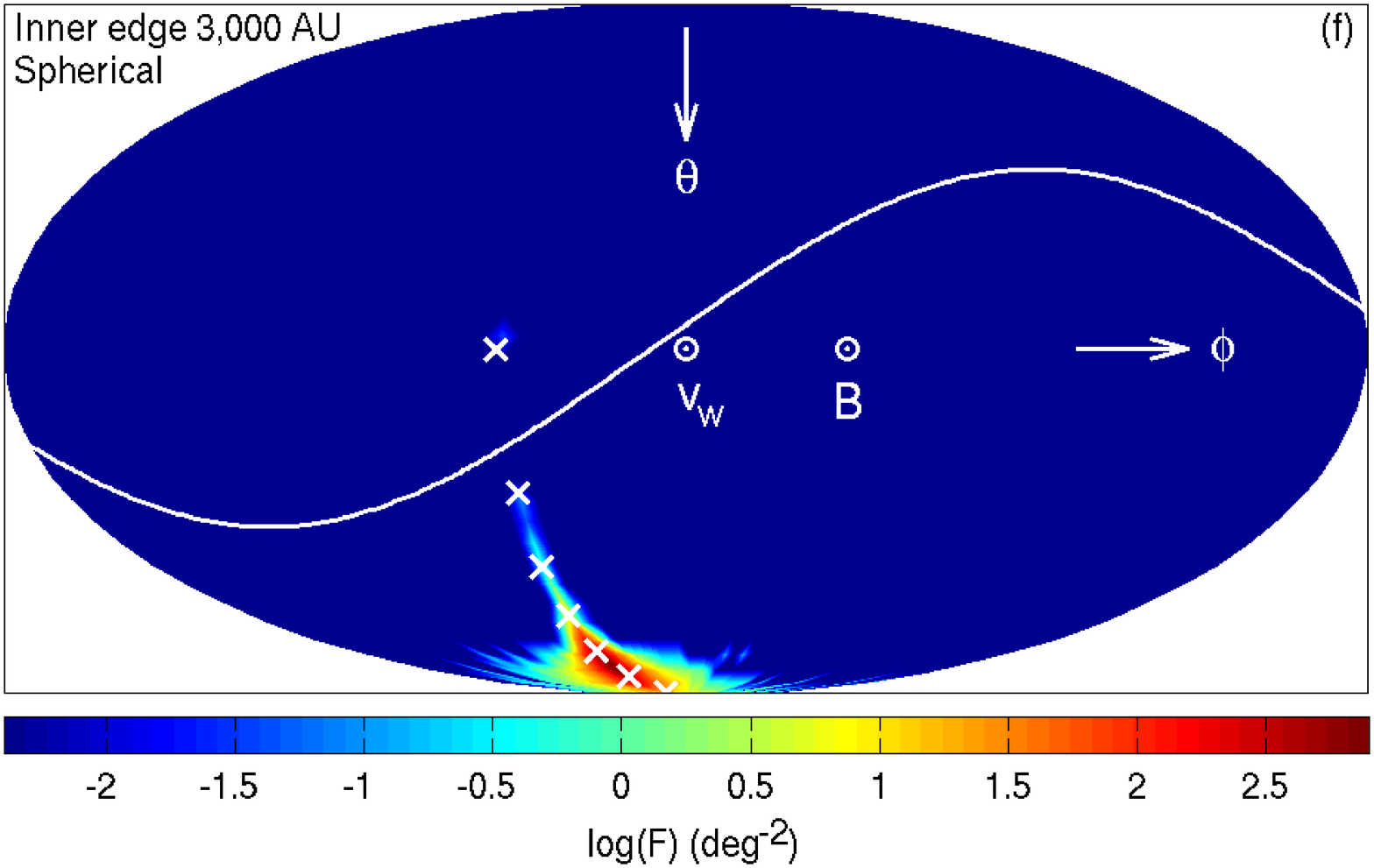}}
\end{center}
\caption{Test of the ``slab'' approach using maps of particle arrival direction, analogous to Figure \ref{map4}. In panels (a)-(d), the particles were created in a slab (of different thickness and for different $r_{in}$, see the labels) perpendicular to the magnetic field direction, while the entire Cloud was populated in panels (e) and (f). The results for a slab of 3,000 AU full thickness are similar to the spherical distribution for both values of $r_{in}$, while those for a slab of $10^3$ AU full thickness show differences, specifically, the confinement of the arrival directions toward the plane perpendicular to the magnetic field. We adopt the 3,000 AU slab thickness for our models.}
\label{mapslab}
\end{figure*}

Particle motion was integrated with a 5th-order Cash-Karp Runge-Kutta method \citep{NumRec} with adaptive step size control. The integration scheme was tested to integrate Keplerian orbits accurately: after 30 dynamical timescales, the relative error in the perihelion position of highly elliptical orbits (of most interest to us) was measured to be no more 0.05\%, and the relative error in perihelion velocity was  $\lesssim 0.02\%$.

A particle was considered to have escaped from the Solar System if it reached $10^5$ AU ($\sim 0.5$ pc) from the Sun. The particle was considered destroyed if it approached within $R_\odot$ from the center of the potential. This also prevented the step size from becoming unmanageably small. Otherwise, the integration was continued for 30 dynamical times, and the particle was considered bound to the Sun.

Given that particles start with velocity $v_0\ll v_w$, their subsequent motion is approximately perpendicular to the magnetic field direction, see \S \ref{sect:small_dyn} and equation (\ref{eq:r_t}). We use this property to speed up the simulation: instead of populating with particles the full spherical extent of the Oort Cloud, we do this only in a thick circular slab (or short cylinder) perpendicular to the $B$-field and passing through the Sun. This rids us of the need to track the motion of particles at the Oort Cloud periphery, which never end up approaching the Sun. We run several test simulations with different widths of the slab as well as with a spherical distribution of particles to confirm that our ``slab'' approach does not introduce any biases. The results are shown in Figure \ref{mapslab}, where we display the maps of the arrival directions of particles produced according to the procedure outlined in \S \ref{vel_dir} for different slab widths in a simulation with $\dot Q(r_0)\propto r_0^{-4}$, $r_{in}=10^3$ AU or $3\times 10^3$ AU, and $\theta_{wB} = 42^\circ$. One can see that the simulation with the 3,000 AU wide slab reliably captures all the details of the particle arrival direction distribution, and we adopt this value in our work.

It should be noted that in our simulations, we always fix a total number of particles {\it inside the slab} $N_{\rm slab}=10^{10}$, which is not equal to the full number of particles $N_{\rm sphere}$ contained in the spherical Cloud populated according to the same radial production profile. For a slab semi-thickness $z$ ($1,500$ AU in our case) and particle production rate $\dot Q(r_0)\propto r_0^{-\chi}$ with $\chi>3$ the two are related via 
\ba
\frac{N_{\rm slab}}{N_{\rm sphere}} =
\left\{
     \begin{array}{lr}
       \frac{\chi-3}{\chi-2}\frac{z}{r_{in}}, ~~~~~~~~~~z<r_{in}\\
       1-\frac{1}{\chi-2}\left(\frac{r_{in}}{z}\right)^{\chi-3}, ~~~z>r_{in}
     \end{array}
   \right.
\label{eq:conversion}
\ea
We always apply this ``slab'' correction when discussing the fraction of cometary grains passing through the ISS.

 
\section{Erosion timescale.}
\label{sect:erosion}


We calculate the erosion timescale $t_{er}$ for a population 
of comets with a number density (per unit volume and unit 
mass) 
\ba
\frac{dn}{dm}=(2-p)\frac{\rho(r)}{m_m^2}
\left(\frac{m}{m_m}\right)^{-p},~~~m<m_m,
\label{eq:com_size_deist}
\ea
where $\rho(r)$ is the total mass density of comets (which is 
a function of time) at distance $r$, $m_m=(4\pi/3)\rho d_m^3$ 
is the characteristic mass of the biggest comets of radius $d_m$, 
and $p<2$ so that the mass of the cometary population is dominated 
by large objects. We define $t_{er}$ as the time during which 
an object with radius $d_m$ (the characteristic 
size at which most of the cometary mass is concentrated) 
experiences a destructive collision. If $m_{pr}(m)$ is the minimum 
mass of a projectile that destroys an object of mass $m$, then the
number density of objects leading to destructive collisions with 
mass $m_m$ is $n(>m_{pr}(m_m))=\int_{m_{pr}(m_m)}^{m_m}
\left(dn/dm\right)dm$, and the erosion time is given by
\ba
t_{er}^{-1}\approx \pi d_m^2 v_r n(>m_{pr}(m_m))
\approx \pi\frac{2-p}{p-1} d_m^2 v_r\frac{\rho(r)}{m_m}
\left[\frac{m_m}{m_{pr}(m_m)}\right]^{p-1},
\label{eq:er}
\ea
where $v_r$ is the characteristic relative velocity between the 
colliding comets, and we used equation (\ref{eq:com_size_deist}) 
and assumed that $m_{pr}(m_m)\ll m_m$.

The minimum projectile mass $m_{pr}(m)$ is set by the condition 
$(1/2)m_{pr}(m)v_r^2=m Q_{\rm D}^\star(m)$, where 
$Q_{\rm D}^\star(m)$ is the specific energy needed for 
catastrophic disruption of a body of mass $m$. We adopt the 
following behavior of $Q_{\rm D}^\star(d)$ as a function of 
object radius $d$, which is consistent with \citet{Benz} 
and \citet{Stewart09} for collisions at speeds not 
exceeding 1 km s$^{-1}$:
\ba
Q_{\rm D}^\star(d)=1.3\times 10^6 \left(\frac{d}{\rm cm}\right)^{-0.4}
+ 0.08 \left(\frac{d}{\rm cm}\right)^{1.3}\mbox{erg g}^{-1}.
\label{eq:Q_D}
\ea
The first term is important for small bodies ($d\lesssim 0.1$ km), 
in the strength-dominated regime, while the second term governs
$Q_{\rm D}^\star$ behavior for large objects ($d\gtrsim 0.1$ km), 
in the gravity-dominated regime. In the following, we will assume a 
simple power law scaling of $Q_{\rm D}^\star(d)$ depending on 
the object size: $Q_{\rm D}^\star(m)=Q_{\rm D,0}^{\star}d^{q}$,
where $Q_{\rm D,0}^{\star}=1.3\times 10^6$ erg g$^{-1}$, $q=-0.4$ 
for $d\lesssim 0.1$ km, and $Q_{\rm D,0}^{\star}=0.08$ erg g$^{-1}$, 
$q=1.3$ for $d\gtrsim 0.1$ km. In the limit $d_{pr}(d)\ll d$, the 
catastrophic disruption condition then yields the minimum 
projectile size ($d_{pr}(d)=d\left[2Q_{\rm D}^\star(d)/v_r^2\right]^{1/3}$) 
of $d_{pr}(d)\approx 2$ m $(d/0.1\mbox{km})^{0.87}r_3^{1/3}$ for 
$d\lesssim 0.1$ km and $d_{pr}(d)\approx 1$ m $(d/0.1 \mbox{km})^{1.43}
r_3^{1/3}$ for $d\gtrsim 0.1$ km, where we set $v_r=v_K(r)$, see 
equation (\ref{eq:v_K}).

\citet{OBrien} demonstrate that for a power law scaling of 
catastrophic disruption strength $Q_{\rm D}^\star(d)\propto d^q$, the 
slope of the fragmentation cascade $p$ is uniquely related to $q$
via $p=(11+q)/(6+q)$. This results in $p\approx 1.89$ and $1.69$
in the strength- and gravity-dominated regimes, respectively.

Substituting equations (\ref{eq:v_K}), (\ref{eq:rho_0}), the expression 
for $d_{pr}$, and the power law scaling of $Q_{\rm D}^\star(d)$ into 
equation (\ref{eq:er}), we find the erosion time as a function of 
distance $r_0$ within the Cloud
\ba
t_{er}(r_0)\approx \frac{16\pi (p-1)}{3(2-p)(\eta-3)} 
\frac{\rho r_{in}^3 d_m^{1+q(p-1)}\left(2Q_{D,0}^\star\right)^{p-1}}
{v_r^{2p-1}(r_{in})M_{\rm OC}}\left(\frac{r_0}{r_{in}}\right)^{\eta+p-1/2}. 
\label{eq:t_er}
\ea
For the two limits following from equation (\ref{eq:Q_D}) with $\eta = 3.5$ we find
\ba
t_{er}(r_0) & \approx & 5~\mbox{Gyr}~d_{m,1}^{0.64}r_{in,3}^{4.39}
M_{\rm OC,10}^{-1}\left(\frac{r_0}{r_{in}}\right)^{4.89},
~~~d_m\lesssim 0.1\mbox{km},
\label{eq:t_er_small}\\
t_{er}(r_0) & \approx & 150~\mbox{Gyr}~d_{m,1}^{1.9}r_{in,3}^{4.19}
M_{\rm OC,10}^{-1}\left(\frac{r_0}{r_{in}}\right)^{4.69},
~~~d_m\gtrsim 0.1\mbox{km}.
\label{eq:t_er_big}
\ea
These estimates suggest that if comets are rather weak aggregates, 
which would make them more compatible with the estimate 
(\ref{eq:t_er_small}), and the bulk of mass in the Oort Cloud is 
concentrated in relatively small objects ($d_m\lesssim 1$ km), 
then the erosion time at the inner edge of the Cloud 
(at $r_0=r_{in}$) can indeed be similar to the Solar 
System age. They also justify our choice of $\kappa=4.5$ in 
equation (\ref{eq:ter_r}).

\end{document}